\documentclass[onecolumn,amsmath,amssymb,11pt,nofootinbib]{revtex4}

\def\gsim{ \lower .75ex \hbox{$\sim$} \llap{\raise .27ex \hbox{$>$}} }

\def\lsim{ \lower .75ex \hbox{$\sim$} \llap{\raise .27ex \hbox{$<$}} }

\usepackage{graphicx}
\usepackage{dcolumn}
\usepackage{bm}
\input epsf
\def\gsim{ \lower .75ex \hbox{$\sim$} \llap{\raise .27ex \hbox{$>$}} }
\def\lsim{ \lower .75ex \hbox{$\sim$} \llap{\raise .27ex \hbox{$<$}} }

\usepackage{graphicx}
\usepackage{dcolumn}
\usepackage{bm}

\newcommand{\nn}{\nonumber}
\newcommand{\be}{\begin{equation}}
\newcommand{\ee}{\end{equation}}
\newcommand{\bea}{\begin{eqnarray}}
\newcommand{\eea}{\end{eqnarray}}

\newcommand{\ie}{{\it i.e.}~}

\def\a{\alpha}

\def\d{\delta}

\def\s{\sigma}

\def\e{\epsilon}

\def\f{\phi}

\def\k{\kappa}

\def\m{\mu}
\def\n{\nu}

\def\th{\theta}

\def\s{\sigma}

\def\t{\tau}

\def\z{\zeta}

\def\pt{\partial}

\begin{document}

\title{Ekpyrotic Non-Gaussianity -- A Review}

\author{Jean-Luc Lehners}
\affiliation{Princeton Center for Theoretical Science, Princeton University, Princeton, NJ 08544 USA \\
}
\email{jlehners@princeton.edu}

\begin{abstract}
Ekpyrotic models and their cyclic extensions solve the standard
cosmological flatness, horizon and homogeneity puzzles by
postulating a slowly contracting phase of the universe prior to
the big bang. This ekpyrotic phase also manages to produce a
nearly scale-invariant spectrum of scalar density fluctuations,
but, crucially, with significant non-gaussian corrections. In
fact, some versions of ekpyrosis are on the borderline of being
ruled out by observations, while, interestingly, the best-motivated models
predict levels of non-gaussianity that will be measurable by
near-future experiments. Here, we review these predictions in
detail, and comment on their implications.
\end{abstract}


\maketitle
\vspace{-1cm}
\tableofcontents{}

\section{Motivation and Introduction}
\label{section intro}

The standard big bang cosmology is hugely successful in
describing the evolution of our universe from the time of
nucleosynthesis onwards. However, a central assumption is that
the universe started out in a hot big bang, and in a special
state: extrapolating back from current knowledge, we know that
early on the universe must have been very flat, homogeneous and
isotropic, with in addition small density perturbations with a
nearly scale-invariant spectrum and a nearly gaussian
distribution. Hence, the ``initial'' state of the universe was
far from random, and its specialness prompts us to try and
explain it via a dynamical mechanism.

The most studied such mechanism is the model of inflation,
which assumes that there was a phase of rapid, accelerated
expansion preceding the hot big bang; for a comprehensive
review see \cite{Baumann:2009ds}. Such a phase can be modeled
by having a scalar field (the ``inflaton'') with a positive and
suitably flat potential. Inflation has the property of
flattening the universe, so that, if it lasts long enough, the
flatness of the ``initial'' state can be explained. Moreover,
inflation possesses the remarkable byproduct that it generates
nearly scale-invariant spectra of scalar and tensor
perturbations by amplifying quantum fluctuations. The predicted
scalar perturbations are in excellent agreement with current
observations, but the tensor perturbations have yet to be
observed - their discovery would be a strong indication for the
correctness of the inflationary picture. However, inflation
also presents a number of conceptual problems: for example,
even though the inflationary phase is supposed to erase all
memory of initial conditions, this is not really the case. In
order for inflation to start in a given patch of space, that
patch must be reasonably smooth over several Planck lengths and
the inflaton field must have a small initial velocity (the
``patch'' and ``overshoot'' problems respectively, see {\it e.g.} \cite{Baumann:2009ds,Brustein:1992nk}). Also, it
has been realized not long ago that inflation is geodesically
incomplete towards the past, which means that the predictions
of the theory depend on the specification of data on a
spacelike initial hypersurface \cite{Borde:2001nh}. In other
words, inflation requires its own initial conditions. Hence, if
inflation is correct, it will only form a part of the story.
More worrying is the problem of unpredictability, which is
associated with the quantum nature and the effectiveness of
inflation. Inflation ends when the inflaton field oscillates
around a minimum of its potential, and ``reheats'' the universe
by decaying into standard model particles. However, for generic initial conditions there will always be regions in which rare but large
quantum fluctuations kick the inflaton field back up
its potential and keep a fraction of the universe in the
inflationary phase. In most of the concrete realizations of
inflation, the region that keeps inflating expands so fast that
it quickly dominates the overall volume of the universe. Hence,
inflation never ends and the global picture of this process of
``eternal inflation'' is that of an empty de Sitter universe
punctured by an infinite number of small pockets where
inflation has ended (at a random time) \cite{Guth:2007ng}.
Because inflation ends at a random moment in these pocket
universes, the pockets might have become sufficiently flattened
or not, they might have acquired scale-invariant perturbations
or not. Without a measure which would determine the relative
likelihood of the various pockets, it becomes difficult to know
exactly what eternal inflation predicts! These problems do not
mean that the idea of inflation is wrong, but, if inflation
continues to be supported by observations, they will have to be
addressed. In the meantime, the seriousness of these open
problems means that it is worthwhile considering alternative
models for the early universe in parallel.

The present review deals with one such model in particular,
namely the ekpyrotic model and its extension, the cyclic
universe; for a comprehensive overview see
\cite{Lehners:2008vx}. In this model, the inflationary phase is
replaced by the ekpyrotic phase, which is a slowly contracting
phase preceding the big bang. The ekpyrotic phase can be
modeled by having a scalar field with a negative and steep
potential. As described in detail below, it also manages to
flatten a given region of the universe, and generates nearly
scale-invariant scalar perturbations, but no observable tensor
fluctuations. At the linear level, the scalar fluctuations are
virtually indistinguishable from the perturbations produced by
inflation, but at higher orders the predictions differ. Since
primordial gravitational waves might turn out to be rather
elusive to measure over the coming years, the most promising
way of distinguishing between alternative models of the early
universe is therefore by studying these higher-order,
non-gaussian signatures.

There is a simply, intuitive argument for why the predictions
regarding higher-order corrections to the linear perturbations
should differ for models of inflation and ekpyrosis. For a
scalar field fluctuation $\delta \varphi,$ the semi-classical
probability density is roughly given by $e^{-S_E(\delta
\varphi)},$ where $S_E(\delta \varphi)$ is the euclidean action
\cite{Seery:2005gb}. Since inflation requires a very flat
potential, the inflaton is an almost free field. For a free
field, the action is quadratic in the field, and hence the
probability distribution is simply a gaussian distribution. For
an exact gaussian distribution the 3-point function $\langle
\delta \varphi^3 \rangle$ vanishes, and hence for inflation,
where the field is almost free, we would expect the 3-point
function to be non-zero, but small. For ekpyrosis, on the other
hand, the potential is steep, and hence the scalar field is
necessarily significantly self-coupled. This has the
consequence that ekpyrotic models generally predict significant
levels of non-gaussianity. In fact, some versions of ekpyrosis
are already on the borderline of being ruled out by
observations, while the best-motivated models predict values
that are measurable by near-future experiments. Thus, the
non-gaussian predictions are crucial in assessing the viability
of various cosmological models, and promise to significantly
enhance our understanding of the physics of the early universe.

The plan of this review is to start with a brief summary of the
main ideas behind ekpyrotic and cyclic models of the universe.
We will then discuss in some detail the generation of linear
cosmological perturbations (a good understanding of the linear
perturbations greatly facilitates an understanding of the
higher-order ones), before turning to the main subject of the
review, namely the non-gaussian corrections to these linear
perturbations. We will conclude with a discussion of the
non-gaussian predictions and in particular their observability
and relation to current observational limits, as well as the
consequences of a potential detection.

\section{Ekpyrotic and Cyclic Cosmology}

The ekpyrotic phase is the cornerstone of ekpyrotic and cyclic
models of the universe: it is a conjectured, slowly contracting
phase preceding the big bang, and it resolves the standard
cosmological puzzles \cite{Khoury:2001wf,Erickson:2006wc}. The
main feature of ekpyrosis is that during this phase the
equation of state \be w \equiv \frac{p}{\rho}\gg 1 \ee is very
large (here $p$ and $\rho$ denote the average pressure and
energy density of the universe). Let us briefly explore the
most direct consequences of such an ultra-stiff equation of
state. Consider a Friedmann-Robertson-Walker (FRW)
metric\footnote{I will mostly use natural units $\hbar=c=1$ and
$8\pi G = M_{Pl}^{-2}=1.$} \be d s^2=-d t^2+a(t)^2\left(\frac{d
r^2}{1-\kappa r^2}+r^2 d \Omega_2^2 \right)
\label{FRWmetric}\ee where $a(t)$ denotes the scale factor of
the universe and $\kappa=-1,0,1$ for an open, flat or closed
universe respectively. If the universe is filled with a number
of fluids interacting only via gravity and with energy densities $\rho_i$ and constant equations of state
$w_i,$ then the equations of continuity  \be \dot{\rho_i} + 3
\frac{\dot{a}}{a}(\rho_i+p_i)=0 \label{continuity}\ee (where
dots denote derivatives with respect to time t) imply that they
will evolve according to \be \rho_i \propto a^{-3(1+w_i)}.
\label{fluidscaling}\ee The Einstein equations for this system
contain a constraint equation, better known as the Friedmann
equation, which involves the Hubble parameter $H\equiv \dot a /
a$: \be H^2 = \frac{1}{3} \left(\frac{-3\kappa}{a^2}+
\frac{\rho_{m,0}}{a^3} + \frac{\rho_{r,0}}{a^4}+ \frac{\rho_{a,0}}{a^6} +
\ldots + \frac{\rho_{\phi ,0}}{a^{3(1+w_{\phi})}} \right).
\label{Friedmann} \ee The $\rho_{i,0}$s are constants giving the energy densities at scale factor $a=1$ of the various constituents of the universe: we
consider the universe to be composed of non-relativistic matter
(subscript $m$), radiation ($r$) and the energy density
associated with anisotropies in the curvature of the universe
($a$). In addition, we consider there to be ekpyrotic (scalar) matter,
denoted by the subscript $\phi$, and, as usual, there is a
contribution due to the average curvature of space.

\begin{figure}[t]
\begin{center}
\includegraphics[width=0.65\textwidth]{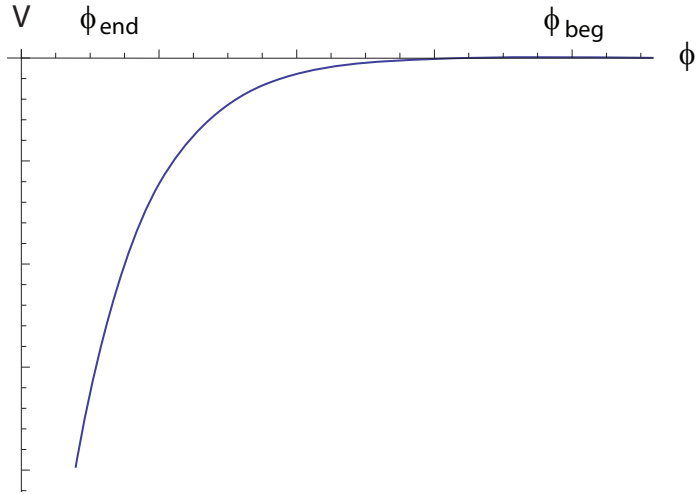}
\caption{\label{figure ekpotential} {\small The potential during
ekpyrosis is negative and steeply falling; it can be modeled by the exponential form $V(\phi)=-V_0 e^{-c\phi}.$}}
\end{center}
\end{figure}

As the universe contracts, components whose energy density
scales with a higher negative power of the scale factor $a$
will successively come to dominate, first matter, then
radiation, then anisotropies and eventually, since $w_\phi \gg
1$ by assumption, the ekpyrotic matter. This means that the
relative energy densities in curvature and anisotropies, for
example, become smaller and smaller, the longer the ekpyrotic
contracting phase lasts. In other words, if ekpyrosis lasts
long enough, the flatness problem is solved. We will make this
statement quantitative below. Strictly speaking, for the
flatness problem to be solved, all we need is a matter
component with $w>1.$ In the next section, we will see that for
realistic ekpyrotic models, typically $w_\phi \gg 1.$ In
passing, we should also point out that there is no horizon
problem in ekpyrotic and cyclic models, as there is plenty of
time before the big bang for different parts of our currently
observable universe to have been in causal contact with each
other.

But what form of matter can have the large equation of
state that we require? A simple way to model the ekpyrotic matter is to have a
scalar field $\phi$ with a steep and negative potential
$V(\phi).$ A concrete example is provided by the negative
exponential \be V(\phi)=-V_0 e^{-c \phi},
\label{ekpotential}\ee where $V_0$ and $c$ are constants - see
Fig. \ref{figure ekpotential}. In the context of string theory,
such scalar fields appear very naturally, and the ekpyrotic
potential can then correspond to an attractive force between
branes - this picture will be briefly described below.

Given an explicit form of the potential, such as
(\ref{ekpotential}), we can solve for the evolution of the
universe. In fact it is straightforward to generalize the
treatment to having many scalars $\phi_i$ with potentials
$V_i(\phi_i).$ Then, in a flat FRW background and neglecting
other matter components, the equations of motion become \be
\ddot{\phi}_i + 3H\dot{\phi}_i + V_{i,\phi_i} = 0
\label{Eomscalar}\ee
 and \be H^2 =\frac{1}{3} \left[\frac{1}{2}  \sum_i
 \dot\phi_i^{~2}+ \sum_i V_i(\phi_i)
 \right], \label{EomFriedmann}\ee
where $V_{i,\phi_i} = (\partial V_i/\partial \phi_i)$ with no
summation implied. If all
the fields have negative exponential potentials
$V_i(\phi_i)=-V_i\, e^{-c_i \phi_i}$ and if $c_i \gg 1$ for all
$i,$ then the Einstein-scalar equations admit the {\it scaling
solution} \be a = (-t)^{1/\e}, \qquad \phi_i = {2\over c_i} \ln
(-\sqrt{c_i^2V_i/2} t), \qquad \frac{1}{\e}=\sum_i {2 \over
c_i^2}. \label{ekpyrosis-scaling} \ee Thus, we have a very
slowly contracting universe with (constant) equation of state
\be w \equiv \frac{  \sum_i \frac{1}{2} \dot\phi_i^{~2}-
V_i(\phi_i)}{\sum_j \frac{1}{2} \dot\phi_j^2+
V_j(\phi_j)}=\frac{2\e}{3}-1 \gg 1.
\label{ekpyrosis-eq-of-state}\ee We are using a coordinate
system in which the big crunch occurs at $t=0;$ in other words,
the time coordinate is negative during the ekpyrotic phase.
Here, the parameter $\e$ corresponds to the {\it fast-roll}
parameter and is typically of ${\cal O}(100)$; its definition
is identical with that in inflation, where its value is
typically of ${\cal O}(1/100)$ and where, correspondingly, it is called the {\it
slow-roll} parameter.

Using this explicit solution, we can get an idea for how long
the ekpyrotic phase has to last in order for the flatness
problem to be solved. Quantitatively, the problem can be
formulated as follows: dividing the Friedmann equation (\ref{Friedmann}) by $H^2$
we can see that the fractional energy density stored in the
average curvature of the universe is given by \be
\frac{\kappa}{(aH)^2}.\ee At the present time, observations
imply that this quantity is smaller than $10^{-2}$ in
magnitude \cite{Komatsu:2008hk}. If we assume a radiation-dominated universe, which
is a good approximation for this calculation, then $aH \propto
t^{-1/2}$ and hence, if we extrapolate back to the Planck time,
the fractional energy density in curvature must have been
smaller than \be \frac{t_{Pl}}{t_0}10^{-2} \approx 10^{-62},
\ee an incredibly small number. However, from
(\ref{ekpyrosis-scaling}), we can see that during the ekpyrotic
phase the scale factor $a$ remains almost constant, while the
Hubble parameter $H\propto t^{-1}.$ Hence $aH$ grows by a
factor of $10^{30}$ as long as \be |t_{ek-beg}| \geqslant
e^{60} |t_{ek-end}|, \label{flatness-constraint-time} \ee where
the subscripts $ek-beg$ and $ek-end$ refer to the beginning and
the end of the ekpyrotic phase respectively. As will be
discussed in the next section, we need $t_{ek-end} \approx
-10^3 M_{Pl}^{-1}$ in order to obtain the observed amplitude of
cosmological perturbations, so that we need \be |t_{ek-beg}|
\geqslant 10^{33} M_{Pl}^{-1} = 10^{-10} s. \ee This is the
minimum time the ekpyrotic phase has to last in order to solve
the flatness problem. Cosmologically speaking, this is a very
short time, attesting to the effectiveness of the ekpyrotic
phase.

Before discussing the cosmological perturbations produced
during the ekpyrotic phase, it is useful to provide a quick
overview of how the ekpyrotic phase might fit into a more
complete cosmological model. The crucial ingredient in any such
model is the proposed mechanism for how the ekpyrotic
contracting phase (with $H<0$) and the subsequent
radiation-dominated expanding phases (with $H>0$) should link
up. The Einstein equations provide the relation \be \dot H =
-\frac{1}{2} (\rho + p). \ee All forms of matter that are
currently known to exist obey the {\it null energy condition}
\be \rho + p \geqslant 0 \qquad {\text{(NEC)}},\ee which
implies $\dot H \leqslant 0$ and which thus precludes a smooth
transition between a contracting and an expanding universe.
This leaves two possibilities for achieving such a transition:
either the NEC is violated during the transition, or the
transition is classically singular.

In {\it new ekpyrotic} models
\cite{Buchbinder:2007ad,Buchbinder:2007tw,Creminelli:2007aq}, a
smooth reversal from contraction to expansion is achieved by
adding a further matter component to the universe which can
violate the NEC. The particular example that these models
consider is the so-called ghost condensate, which corresponds
to the gravitational equivalent of a Higgs phase
\cite{ArkaniHamed:2003uy}. It is not clear yet whether or not
the ghost condensate can be obtained from a fundamental theory
such as string theory \cite{Adams:2006sv} (in the more restricted framework of quantum field theory it seems impossible to construct a stable ghost condensate model \cite{Kallosh:2007ad}); however, it is interesting that string
theory contains many objects (orientifolds, negative-tension
branes) which do violate the NEC. Of course, simply adding such
a component is not enough: it must become relevant as the
universe contracts, and vanish again as the universe expands.
The simplest way in which to achieve this is by assuming that
the ghost condensate itself also plays the role of the
ekpyrotic matter, and that after the transition to expansion,
it decays into ordinary matter fields. This scenario requires
the ghost condensate to possess both a particular form for its
kinetic term and a particular potential; for details regarding
possible realizations see
\cite{Buchbinder:2007ad,Creminelli:2007aq}.

\begin{figure}[t]
\begin{center}
\includegraphics[width=0.65\textwidth]{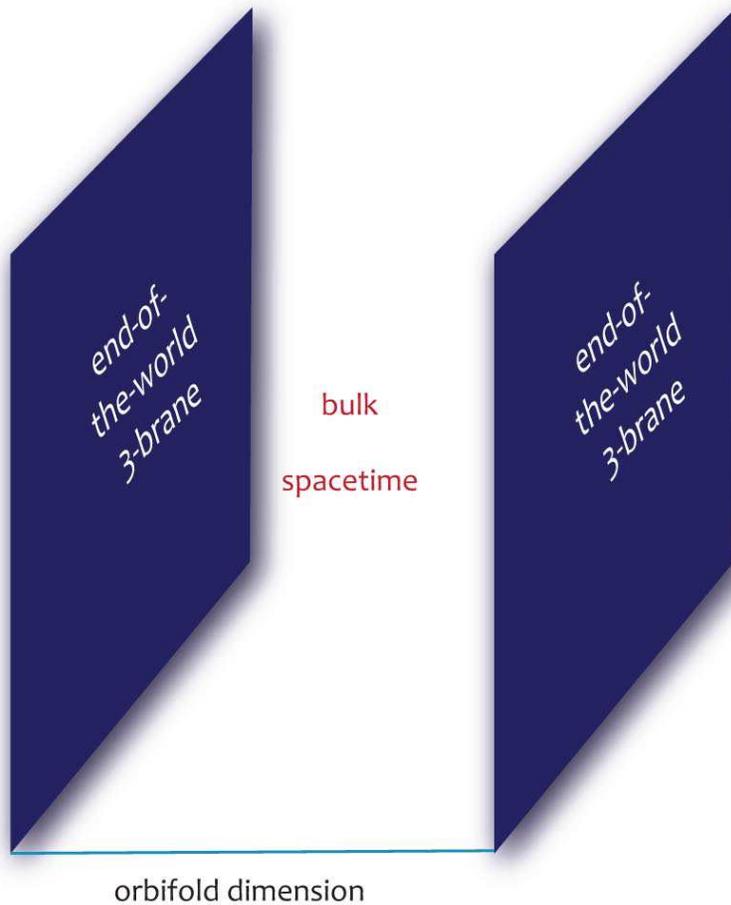}
\caption{\label{figure braneworld} {\small The braneworld picture
of our universe. Think of a sandwich: the
5-dimensinonal bulk spacetime is bounded by two 4-dimensional boundary branes. There is no
space ``outside'' of the sandwich, but the branes can be infinite
in all directions perpendicular to the line segment (orbifold). In the M-theory embedding, there are 6 additional internal dimensions at each point of the sandwich.}}
\end{center}
\end{figure}

The {\it cyclic} model of the universe
\cite{Steinhardt:2001vw,Steinhardt:2001st} is based on the
braneworld picture of the universe, in which spacetime is
effectively 5-dimensional, but with one dimension not extending
indefinitely, but being a line segment, see Fig. \ref{figure
braneworld}. The endpoints of this line segment (orbifold) are
two $(3+1)$-dimensional boundary branes. In the full string
theory setup, there is in addition a 6-dimensional internal
manifold at each point in the 5-dimensional spacetime, for a
total of 11 dimensions \cite{Lukas:1998yy}. This description of
the universe stems from string theory, and in particular the
duality, known as {\it Ho\v{r}ava-Witten theory}
\cite{Horava:1995qa}, between 11-dimensional supergravity and
the $E_8 \times E_8$ heterotic string theory. All matter and
forces, except for gravity, are localized on the branes, while
gravity can propagate in the whole spacetime. Our universe, as
we see it, is identified with one of the boundary branes and,
as long as the branes are far apart, can interact with the
other brane only via gravity. The cyclic model assumes that
there is an attractive force between the two branes, which
causes the branes to approach each other. This force is modeled
by a potential of the form shown in Fig. \ref{figure
cycpotential}. Note that the potential incorporates an
ekpyrotic part. From the higher-dimensional point of view, the
ekpyrotic phase has the rather non-intuitive property that it
flattens the branes to a very high degree. Eventually the two
branes collide and bounce off each other. It is this collision
that, from the point of view of someone living on one of the
branes, looks like the big bang. Classically, the collision is
singular, as the orbifold dimension shrinks to zero size. The
collision is slightly inelastic and produces matter and
radiation on the branes, where the standard cosmological
evolution now takes place. However, due to quantum
fluctuations, the branes are slightly rippled and do not
collide everywhere at exactly the same time. In some places,
the branes collide slightly earlier, which means that the
universe has a little bit more time to expand and cool. In
other places, the collision takes place slightly later, and
those regions remain a little hotter. This provides a heuristic
picture of the way temperature fluctuations are naturally
produced within the model. Shortly after the branes have
separated, the distance between the boundary branes gets almost
stabilized, but the branes start attracting each other again
very slightly. This very slight attraction acts as
quintessence, and is identified with the dark energy observed
in the universe. After a long time, and as the branes become
closer again, they start attracting each other more strongly so
that we get another ekpyrotic phase and eventually another
brane collision with the creation of new matter. In this way, a
cyclic model of the universe emerges. Before continuing, we
should mention the main open issues related to the cyclic
model: the first one concerns the potential, which at this
point is simply conjectured. It will be important to see if a
potential of the required shape can be derived from
microphysics. And the second is the brane collision, which so
far has been extensively studied at the classical and
semi-classical level \cite{Turok:2004gb}, but a full quantum
treatment has remained elusive.

\begin{figure}[t]
\begin{center}
\includegraphics[width=0.75\textwidth]{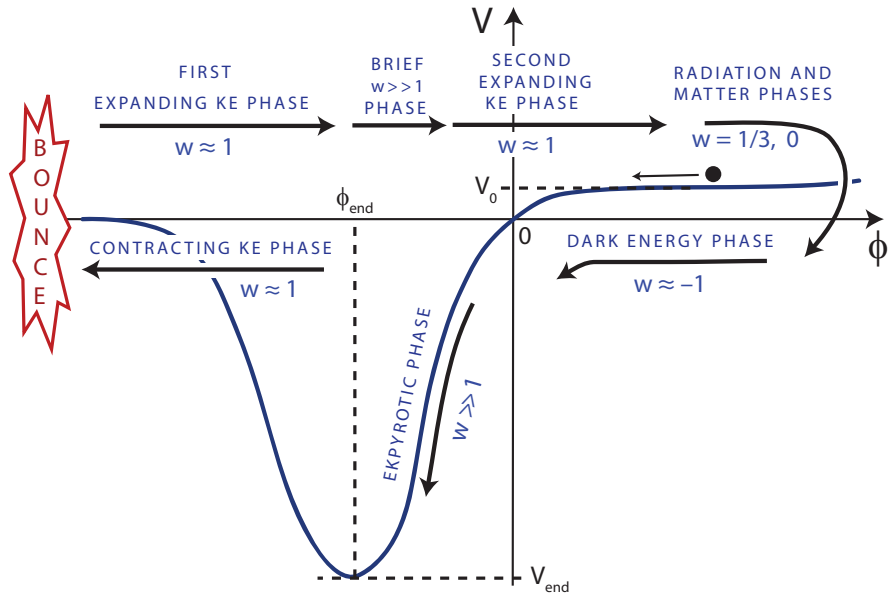}
\caption{\label{figure cycpotential} {\small
The potential for the cyclic universe integrates the ekpyrotic part and a quintessence epoch, but is irrelevant at the brane collision. A possible form for the potential is $V(\f)=V_0(e^{b\f}-e^{-c\f})F(\f),$ with $b\ll 1, \, c\gg 1$ and $F(\f)$ tends to unity for $\f>\f_{end}$ and to zero for $\f < \f_{end}.$ Reproduced with permission from \cite{Erickson:2006wc}.}}
\end{center}
\end{figure}

In the discussion above, we have mostly focussed on models
involving one effective scalar field. However, there are two
good reasons to extend the analysis to two or more scalars:
first, in embedding the ekpyrotic and cyclic models in
M-theory, there are two universal scalars, namely the radion
mode (which determines the distance between the branes) and the
volume modulus of the internal 6-dimensional manifold
\cite{Lehners:2006pu}. There can be many more scalar fields
(such as the shape moduli of the internal space), but we always
must consider these two universal scalars. And
secondly, as we will see in the next section, it is much more
natural to generate a nearly scale-invariant spectrum of
curvature perturbations (in agreement with observations) in
models with two scalars than in models with only one. However,
multi-field ekpyrotic models present some qualitatively new
features, which we discuss briefly here.

\begin{figure}[b]
\begin{center}
\includegraphics[width=0.7\textwidth]{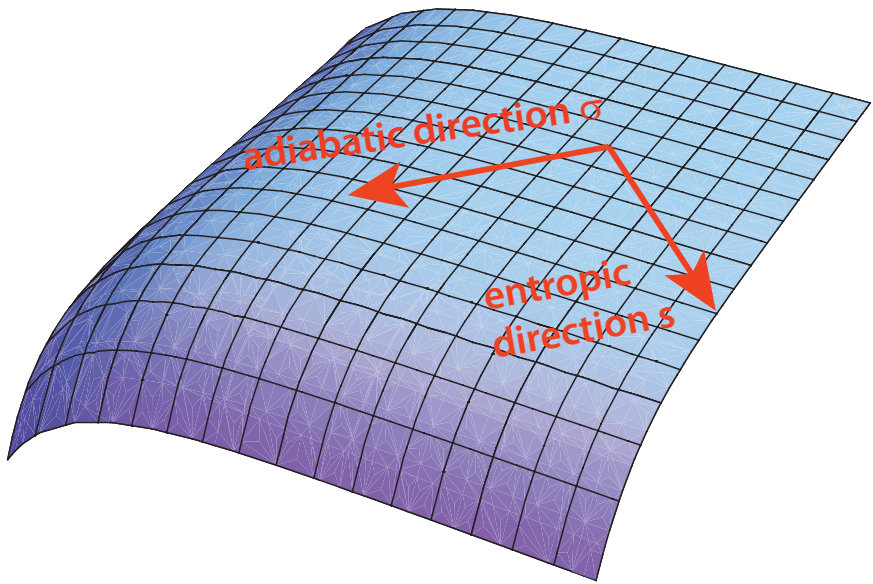}
\caption{\label{figure 2fieldpotential} {\small
After a rotation in field space, the two-field ekpyrotic potential can be viewed as composed of an ekpyrotic direction ($\s$) and a transverse tachyonic direction ($s$). The ekpyrotic scaling solution corresponds to motion along the ridge of the potential. Perturbations along the direction of the trajectory are adiabatic/curvature perturbations, while perturbations transverse to the trajectory are entropy/isocurvature perturbations.}}
\end{center}
\end{figure}

The 4-dimensional effective action \be S=\int
\sqrt{-g}[R-\frac{1}{2}(\pt\phi_1)^2
-\frac{1}{2}(\pt\phi_2)^2-V(\phi_1,\phi_2)]\ee can be obtained
as the low-energy limit of Ho\v{r}ava-Witten theory, where
$\phi_1$ and $\phi_2$ are related by a field redefinition to
the radion and the internal volume modulus
\cite{Lehners:2006ir}. We are assuming that during the
ekpyrotic phase, both fields feel an ekpyrotic-type potential,
{\it e.g.} \be V(\phi_1,\phi_2) =-V_1 e^{-c_1 \phi_1} - V_2
e^{-c_2 \phi_2}. \label{potential2field}\ee  Then it is much
more natural to discuss the dynamics in terms of the new
variables $\s$ and $s$ pointing transverse and perpendicular to
the field velocity respectively
\cite{Koyama:2007mg,Koyama:2007ag}; they are defined, up to
unimportant additive constants which we will fix below, via \be
\s \equiv \frac{\dot\phi_1 \phi_1 + \dot\phi_2 \phi_2}{\dot\s},
\qquad s \equiv \frac{\dot\phi_1 \phi_2 - \dot\phi_2
\phi_1}{\dot\s}, \ee with $\dot\s \equiv (\dot\phi_1^2 +
\dot\phi_2^2)^{1/2}.$ It is also useful to define the angle
$\th$ of the trajectory in field space, via
\cite{Gordon:2000hv} \be \cos \th =\frac{\dot\phi_1}{\dot\s},
\qquad \sin \th = \frac{\dot\phi_2}{\dot\s}. \ee In terms of
these new variables, the potential can be re-expressed as \be
V_{ek}=-V_0 e^{\sqrt{2\e}\s}[1+\e s^2+\frac{\k_3}{3!}\e^{3/2}
s^3+\frac{\k_4}{4!}\e^2 s^4+\cdots],
\label{potentialParameterized}\ee where for exact exponentials
of the form (\ref{potential2field}), one has
$\k_3=2\sqrt{2}(c_1^2-c_2^2)/|c_1 c_2|$ and $\k_4=4(c_1^6 +
c_2^6)/(c_1^2 c_2^2(c_1^2 + c_2^2)).$ However, in the absence
of a microphysical derivation of the potential, we will simply
take $\k_3,\k_4 \sim {\cal O}(1)$ and express all results in
terms of $\k_3,\k_4.$ See also Fig. \ref{figure
2fieldpotential} for an illustration of the potential. The
ekpyrotic scaling solution becomes \be a(t)=(-t)^{1/\e} \qquad
\s=-\sqrt{\frac{2}{\e}}\ln \left(-\sqrt{\e V_0} t\right) \qquad
s=0, \label{ScalingSolution}\ee with the angle $\th$ being
constant. The solution corresponds to motion along a ridge in
the potential, as is evident from the figure. Hence, in
contrast to the single field case, the multi-field ekpyrotic
background evolution is unstable to small perturbations
\cite{Lehners:2007ac,Tolley:2007nq}. This implies that the trajectory must be localized near the ridge with extreme precision at the beginning of the ekpyrotic phase, the condition being that the field should stray no more than a value of $e^{-60}$ (at best) in Planck units from the ridge at the beginning of ekpyrosis \cite{Buchbinder:2007tw}. Thus, at first sight, it looks as if the multi-field ekpyrotic phase has not managed to solve the problem of initial conditions. However, there currently exist two approaches addressing this issue: the authors of \cite{Buchbinder:2007tw} considered the existence of a ``pre-ekpyrotic'' phase during which the potential is curved upwards and during which the trajectory is localized. Meanwhile, in the context of the cyclic universe, there is a natural resolution of the issue of initial conditions, not involving any new ingredients of the model: indeed, the multi-field cyclic universe {\it selects} those regions that happen to correspond to trajectories sufficiently close to the ridge, in the sense that these regions are vastly amplified over the course of one cycle due to the phases of radiation, matter and dark energy domination (note that the ekpyrotic phase shrinks the universe by a negligible amount). At the same time, the regions corresponding to trajectories not sufficiently close to the ridge (this would include the vast majority of trajectories) do not undergo a full ekpyrotic phase, and after these regions undergo chaotic mixmaster behavior close to the big crunch, they simply collapse (presumably they will end up forming black holes) and stop both growing and cycling. In this way the global structure of the universe becomes of the {\it phoenix} type, in which vast habitable regions are interspersed with small collapsed ones. The important point is that the habitable regions, which are the only regions of interest to us here, automatically correspond to the regions that had the right ``initial conditions'' at the beginning of their preceding ekpyrotic phase. This is discussed in detail in \cite{Lehners:2008qe}; see
also the essay \cite{Lehners:2009eg}.

\section{Linear Perturbations} \label{sectionlinear}

\subsection{Single Scalar Field}

In the last section, we have dealt with the classical evolution
during the ekpyrotic phase. We will now add quantum
fluctuations and we will see that, just as in inflation, the
quantum fluctuations get amplified into classical density
perturbations. Hence, on top of resolving the standard
cosmological puzzles, the ekpyrotic phase can also be the
source of the primordial temperature fluctuations whose imprint
is seen in maps of the cosmic microwave background, provided that the amplitude and spectrum of the fluctuations match
observations.

From the study of inflationary models, we have developed the
intuition that quantum fluctuations that get stretched to
super-horizon scales turn into classical perturbations, roughly
speaking because the fluctuations go out of causal contact with
themselves, do not remember locally that they are in fact
fluctuations and end up as local, classical perturbations to
the background evolution. In inflation, this effect occurs
because the horizon is approximately constant in size while the
wavelengths of the quantum modes get stretched exponentially
with time (the scale factor of the universe grows
exponentially). For ekpyrosis, the scaling solution
(\ref{ekpyrosis-scaling}) shows that the scale factor is almost
constant, so that the mode wavelengths remain almost constant
too. However, the horizon, which is proportional to $1/H \sim
t,$ shrinks rapidly as $t \rightarrow 0$ and hence the modes
automatically become of super-horizon size\footnote{Since
tensor modes/gravitational waves depend on the evolution of the
scale factor alone, and since the scale factor shrinks
imperceptively slowly during ekpyrosis, there are no
substantial gravity waves produced during the ekpyrotic phase
(the background spacetime is almost Minkowski!)
\cite{Boyle:2003km}. In fact, the dominant gravitational waves
that are produced from ekpyrosis are those that arise from
the backreaction of the scalar fluctuations onto the metric, at
second order in perturbation theory \cite{Baumann:2007zm}.}. We
will now discuss in some detail what amplitude and spectrum
these modes obtain. We will first concentrate on the single
field case, before discussing two fields.

Since the scale factor evolves very little during the ekpyrotic
phase, one is tempted to simply turn gravity off as a first
approximation, and to consider the theory consisting only of a
scalar field with a steep and negative potential
\cite{Khoury:2001zk}: \be S = \int d^4 x [-\frac{1}{2} (\pt
\phi_1)^2 + V_1 e^{-c_1\phi_1}]. \ee Then, if we define scalar
fluctuations $\d\phi$ via $\phi_1 \equiv \bar\phi_1(t) +
\d\phi(t,\underline{x}),$ where
$\bar\phi_1=\frac{2}{c_1}\ln(-\sqrt{c_1^2V_1/2}t)$ denotes the
background evolution, the equation of motion for the
fluctuations is given by \be
\ddot{\d\phi}-\nabla^2\d\phi+V_{,\phi_1\phi_1}\d\phi=0, \ee
where $V_{,\phi_1\phi_1}=-2/t^2.$ We then expand the
fluctuation field $\d\phi$ into Fourier modes \be \d\phi =
\int \frac{d^3 k}{(2\pi)^3} a_{\bf k} \chi_k e^{i{\bf k}\cdot {\bf x}} +
h.c. \ee where the $\chi_k$s are the positive frequency mode
functions (due to the assumed cosmological symmetries, they
depend only on the magnitude $k=|{\bf k}|$). We proceed to
quantize the field by imposing the canonical commutation
relations \be [a_{{\bf k}},a_{{\bf k'}}]= [a^\dag_{{\bf
k}},a^\dag_{{\bf k'}}]=0, \quad [a_{{\bf k}},a^\dag_{{\bf
k'}}]= (2\pi)^3 \d ({\bf k}-{\bf k'}) \ee In the process, the $a_{{\bf
k}}$s have been promoted to (annihilation) operators, and the
vacuum state $|0\rangle$ is defined by $a_{{\bf
k}}|0\rangle=0.$ The mode functions obey the equation of motion
\be \ddot{\chi_k} + k^2 \chi_k -\frac{2}{t^2} \chi_k = 0, \label{modeequation}\ee
which admits the two solutions $\chi_k\propto
e^{-ikt}(1-\frac{i}{kt}),e^{ikt}(1+\frac{i}{kt}).$ However, as $t\rightarrow - \infty$ the modes should asymptote
to the Minkowski space free particle state $\chi_k \rightarrow
e^{-ikt}/\sqrt{2k}$ (note that in that limit (\ref{modeequation}) reduces to the equation of a simple harmonic oscillator), and this fixes the
solution to be \be \chi_k = \frac{1}{\sqrt{2k}}
e^{-ikt}(1-\frac{i}{kt}). \label{modefct}\ee Towards the end of
the ekpyrotic phase, we have $|kt|\ll 1,$ and then the solution
can be well approximated by \be \chi_k \approx
\frac{-i}{\sqrt{2}k^{3/2}t}. \ee

The quantum fluctuations have a mean that is zero, $\langle
0|\d\phi|0\rangle=0.$ However, the {\it variance}
$\Delta_\phi^2(k),$ which is defined by $\langle
0|\d\phi^2|0\rangle \equiv \int \frac{dk}{k} \Delta_\phi^2(k)$,
does not vanish. It is conventional to write the variance as
\be
\Delta_\phi^2(k)=\Delta_\phi^2(k_0)(\frac{k}{k_0})^{n_s-1},\ee
where $k_0$ denotes a reference scale and $n_s$ is the {\it
spectral index}.

A related concept in momentum space is the {\it power spectrum}
$P(k),$ defined by \be P(k) \equiv |\chi_k|^2 =
\frac{2\pi^2}{k^3}\Delta_\phi^2(k). \label{spectrumvariance}\ee
It is the Fourier transform of the 2-point correlation
function, and we can equivalently define it as \be \langle
\z_{{\bf k}} \z_{{\bf k'}}\rangle \equiv (2\pi)^3 P(k)
\d^3({\bf k} + {\bf k'})\ee where isotropy dictates that $P$
only depends on $k=|{\bf k}|.$ We will find this definition
useful later on.

In our case, we have that at late times \be \langle
0|\d\phi^2|0\rangle = \int \frac{d^3 k}{(2\pi)^3} \chi_k^\ast
\chi_k = \int \frac{4\pi k^2 dk}{(2\pi)^3}\frac{1}{2k^3
t^2},\ee so that the variance is given by \be \Delta_\phi^2(k)
= \frac{1}{4\pi^2 t^2} \qquad \rightarrow \quad n_s=1.
\label{varaincesinglefield}\ee The variance is independent of
$k$, and hence we obtain a scale-invariant spectrum for
$\d\phi.$ This looks very promising! However, we really must
include gravity in our analysis, and calculate the spectrum for
the curvature perturbation $\zeta$, which is the quantity that
is measured to have a nearly scale-invariant spectrum of
perturbations.

Once we add gravity, it is easiest to perform the calculation
in so-called $\z$-gauge, where the perturbations in the scalar
field are gauged away and all perturbations are expressed via
dilatations of the 3-metric: \bea \d\phi &=& 0 \\ ds^2 &=&
-dt^2 + a^2(t) e^{2\z(t,{\bf x})} dx_j dx^j, \eea where
$j=1,2,3.$ Then, using the background scaling solution
(\ref{ekpyrosis-scaling}), the action reduces to an action for
$\z$ which is given by \cite{Creminelli:2004jg} \be S=-\int \e
g^{\m\n} \pt_\m \z \pt_\n \z. \ee During ekpyrosis, $\e$ is
typically nearly constant. In fact, in the scaling solution
used above, we have already made the approximation that $\e$ is
constant, and with this approximation, the equation of motion
for $\z$ resulting from the action above is particulary simple:
in Fourier space it is given by \be \ddot\z_k + 3H\dot\z_k +
\frac{k^2}{a^2} \z_k =0. \ee If we use conformal time $\t$,
defined via $dt \equiv a d\t,$ and the notation $'\equiv
\frac{d}{d\t},$ the above equation becomes \be \z_k''
+2\frac{a'}{a}\z_k' + k^2 \z_k=0.\ee After a further change of
variables to $y\equiv a\z/\sqrt{-k\t}$ and $x\equiv -k\t,$ the
equation turns into a Bessel equation $x^2\frac{d^2
y}{dx^2}+x\frac{dy}{dx}+(x^2-\a^2)y=0,$ with $\a=\sqrt{\t^2
a''/a + 1/4}\approx 1/2$ since $a\approx constant$. Hence the
solutions are given by the Hankel functions $y\propto
H_{1/2}^{(1)}(-k\t),H_{1/2}^{(2)}(-k\t)$ and with the boundary
condition that we want $\z \rightarrow e^{-ik\t}/\sqrt{2k}$ as
$\t \rightarrow -\infty,$ we obtain the solution (up to a phase)\footnote{Useful asymptotic expressions are $H^{(1)}_\a(x) \rightarrow \sqrt{\frac{2}{\pi x}}e^{i(x-\a\pi/2-\pi/4)}$ when $x\gg\a$ and $H^{(1)}_\a(x) \rightarrow -\frac{i}{\pi}\Gamma(\a)(\frac{2}{x})^{\a}$ when $x\ll\a$ and for $\a>0.$} \be \z =
\frac{\sqrt{-\t}}{a}H_{1/2}^{(1)}(-k\t). \ee At late times, the
variance becomes \be \langle 0 | \z^2 | 0 \rangle = \int
\frac{d^3 k}{(2\pi)^3}\frac{(-\t)}{a^2}|H_{1/2}^{(1)}(-k\t)|^2
\sim \int \frac{dk}{k} k^2, \ee and hence we get a spectral index
$n_s =3.$ This spectrum is {\it blue}, as there is more power
on smaller scales, and it is in disagreement with observations
\cite{Khoury:2001zk,Lyth:2001pf,Creminelli:2004jg}. Hence, the
scale-invariant spectrum of the scalar perturbation in the no-gravity theory did not get
transferred to the curvature perturbation $\z.$ A closer
analysis reveals that these two perturbations correspond to two
physically distinct modes, the former being a time-delay mode
to the big crunch, and the latter a local dilatation in space.
In a contracting universe, these two modes are distinct, and
they do not mix. It is conceivable that they might mix at the
big crunch/big bang transition
\cite{Tolley:2003nx,McFadden:2005mq}, in which case the
scale-invariant contribution would be the dominant one on the
large scales of interest, but this possibility is still
insufficiently understood to make definite predictions. As we
will show next, this is also unnecessary, as there is a very
natural {\it entropic} mechanism which generates
scale-invariant curvature perturbations before the big bang, as
long as there is more than one scalar field
present\footnote{Recently, Khoury and Steinhardt have also
pointed out that right at the onset of the single-field
ekpyrotic phase, a range of scale-invariant modes can be
produced \cite{Khoury:2009my}. However, contrary to the cases
that we have discussed so far, this {\it adiabatic} mechanism
requires the universe to already be contracting when the
equation of state is still near $w\approx -1.$ If viable, this
mechanism would produce an interesting non-gaussian signal; but
as it is currently not known how to incorporate this mechanism
into a more complete cosmological model, we will not discuss
this mechanism here. See also \cite{Linde:2009mc} for the challenges that this scenario must address.}.
But before continuing, it might be useful to add a few remarks concerning the validity of our approach: indeed, the reader might be worried about the validity of perturbation theory, since the background quantities, such as the Hubble rate, as well as the perturbations themselves blow up as $t\rightarrow 0.$ However, as shown in \cite{Creminelli:2004jg}, after switching to synchronous gauge, it is straightforward to see that the universe evolves to become closer and closer to the unperturbed background solution, and hence perturbation theory is valid. Also, even though the background quantities blow up as seen from the 4-dimensional viewpoint, in fact in the higher-dimensional colliding branes picture the ekpyrotic phase has the effect of flattening the branes and hence of rendering the curvatures {\it small} as the big crunch is approached \cite{Turok:2004gb}.

\subsection{Two Fields: the Entropic Mechanism}

As discussed at the end of the last section, it is rather
unnatural to consider only a single scalar field in the
effective theory, since there are two universal
scalars that are always present in a higher-dimensional
context: the radion field, determining the distance between the
two end-of-the-world branes, and the volume modulus of the
internal manifold. But as soon as there is more than one
scalar field present, one can have entropy, or isocurvature,
perturbations, which are growing mode perturbations in a
collapsing universe \cite{Finelli:2002we}. Entropy
perturbations can source the curvature perturbation, and hence
(provided the entropy perturbations acquire a nearly
scale-invariant spectrum), nearly scale-invariant curvature
perturbations can be generated just before the bounce
\cite{Notari:2002yc,Lehners:2007ac}. These then turn into
growing mode perturbations in the ensuing expanding phase.

For the two scalar fields, we will again assume the potential
(\ref{potentialParameterized}), but with the slight
generalization that we allow the fast-roll parameter $\e$ to be
slowly varying. There are two gauge-invariant scalar
perturbation modes: the entropy
perturbation $\d s = \cos \th \d\phi_2 - \sin\th \d\phi_1$
corresponds to perturbations transverse to the background
trajectory, see Fig. \ref{figure 2fieldpotential}, while the
adiabatic, or curvature, perturbation $\z$ is the
gauge-invariant quantity expressing perturbations along the
background trajectory, see \cite{Gordon:2000hv,Langlois:2006vv}
for a detailed exposition. For a straight trajectory
($\dot\th=0$), the linearized equation of motion for $\d s$ is
\be \ddot{\delta s} + 3H\dot{\delta s} + \left(\frac{k^2}{a^2}
+ V_{ss} \right) \delta s = 0, \label{eq-entropylinear}\ee
where $V_{ss}$ denotes the second derivative of the potential
w.r.t. $s.$ In conformal time, and for the re-scaled variable
$\d S = a(\t) \d s,$ we obtain \be {\delta S}'' + \left(k^2
-\frac{a''}{a} + a^2 V_{ss} \right) \delta S = 0.
\label{eq-entropy-S}\ee To proceed, we must relate $a''/a$ and
$V_{ss}$ to the fast-roll parameter $\e$ and its derivative
w.r.t. the number of e-folds of expansion $N$, where $dN\equiv
d\ln a.$ By requiring $\e$ to vary slowly, what is meant is
that we will keep terms in $d\e/dN,$ but not higher-order terms
such as $d^2\e/dN^2.$ Then, by differentiating
$\e=\dot\s^2/(2H^2)$ twice, and using
$\ddot\s+3H\dot\s+V_{\s}=0$ as well as $V_{ss}=V_{\s\s},$ one
can derive the following expressions, valid to sub-leading
order in $\e$: \bea \frac{a''}{a} &=& H^2 a^2 (2-\e), \\
V_{ss}&=& H^2 \,(- 2 \epsilon^2  + 6 \epsilon +\frac{5}{2}
\,\epsilon_{,N}).\eea Using in addition that
$aH=(1+1/\e+\e_{,N}/\e^2)/(\e\t),$ Eq. (\ref{eq-entropy-S})
finally reads \be \d S'' + \Big(k^2 -\frac{2(1 - \frac{3}{2
\epsilon} + \frac{3}{4}
\frac{\epsilon_{,N}}{\epsilon^2})}{\t^2} \Big)\d S =0. \ee In
analogy with our discussion of the single-field case, this
equation can be solved in terms of the Hankel functions,
supplemented by the boundary condition of approaching the
Minkowski vacuum state in the far past, to yield (up to a
phase) \be \d S = \frac{\sqrt{-k\t}}{2}H^{(1)}_\n (-k\t),
\qquad \n = \frac{3}{2}(1 - \frac{2}{3 \epsilon}
+\frac{\epsilon_{,N}}{3 \epsilon^2}).\ee At late times $(-k\t)
\rightarrow 0$ and we obtain \be \d S \approx
\frac{1}{\sqrt{2}(-\t)k^\n},\ee implying that at the end of the
ekpyrotic phase, the entropy perturbation is given by \be \d
s(t_{ek-end}) \approx \frac{|\e V_{ek-end}|^{1/2}}{\sqrt{2}k^\n}. \label{entropylineargenerated}\ee
Following the same steps as in the single field case above, it
is straightforward to see that the spectral index of the entropy
perturbation is now given by \cite{Lehners:2007ac} \be
\label{tilt1} n_s -1 = \frac{2}{\epsilon } -
\frac{\epsilon_{,N}}{\epsilon^2}. \ee The first term on the
right-hand side is a gravitational contribution, which, being
positive, tends to make the spectrum blue. The second term is a
non-gravitational contribution, which tends to make the
spectrum red. A simple way to estimate the natural range of
$n_s$ is to rewrite the above expression in terms of ${\cal
N}$, the number of e-folds before the end of the ekpyrotic
phase (where $d{\cal N} = d \ln(aH)$): \be n_s -1 =
\frac{2}{\epsilon} -  \frac{d \ln \epsilon}{d {\cal N}}. \ee In
this expression, $\epsilon({\cal N})$ measures the equation of
state during the ekpyrotic phase, which decreases from a value
much greater than unity to a value of order unity in the last
${\cal N}$ e-folds. If we estimate  $\epsilon \approx {\cal
N}^{\alpha}$ \cite{Khoury:2003vb}, then  the spectral tilt is \be n_s-1 \approx
\frac{2}{{\cal N^{\alpha}}} - \frac{\alpha}{{\cal N}}.
\label{tilt2} \ee Here we see that the sign of the tilt is
sensitive to $\alpha$. For nearly exponential potentials
($\alpha \approx 1$),  the spectral tilt is $n_s \approx 1+
1/{{\cal N}} \approx 1.02$, slightly blue, because the first
term dominates.  However, in the cyclic model the steepness of
the potential must decrease in order for the ekpyrotic phase to
come to an end, and $\alpha$ parameterizes these cases.  If
$\alpha > 1.14$,
 the spectral tilt is red.  For example, $n_s = 0.97$ for
$\alpha \approx 2$.   These examples represent  the range that
can be achieved by the entropic mechanism, roughly \be 0.97 <
n_s < 1.02. \ee These are in good agreement with the present
observational bounds obtained by the WMAP satellite, which are
$n_s=0.96 \pm 0.03$ at the $2\s$ level \cite{Komatsu:2008hk}.

Now that we have shown how an approximately scale-invariant
spectrum of entropy perturbations may be generated by scalar
fields in a contracting universe, we will discuss how these
perturbations may be converted to curvature perturbations.
Since the entropy perturbations of interest are all of
super-horizon scales, we can now restrict our study to large
scales only, where spatial gradients can be neglected. On these
scales, the evolution equation for the curvature perturbation
is given by \cite{Gordon:2000hv} \be \dot\z =
-\frac{2H}{\dot\s}\dot\th \d s = \sqrt{\frac{2}{\e}}\dot\th \d
s. \label{zetalinear}\ee Hence, as soon as the background
trajectory bends ($\dot\th \neq 0$), the entropy perturbations
become a source for the curvature perturbations.

There are at least two ways in which such a bending can occur:
the first makes use of the instability of the two-field
ekpyrotic potential, {\it cf.} again Fig. \ref{figure
2fieldpotential}. If the background trajectory strays
sufficiently far from the ridge of the potential, the
trajectory will turn and fall off one of the steep sides of the
potential \cite{Koyama:2007mg,Koyama:2007ag,Buchbinder:2007ad}.
The turning of the trajectory then immediately results in the
conversion of entropy into curvature perturbations. Since this
conversion occurs during the ekpyrotic phase, we will term this
process {\it ekpyrotic conversion}. It is straightforward to
estimate the amplitude of the resulting curvature perturbation
(its spectrum will be identical to the spectrum of the entropy
perturbations, as Eq. (\ref{zetalinear}) is $k$-independent):
if we approximate the entropy perturbation as remaining
constant during the conversion process, and assume a total
bending angle of order unity, $\int \dot \th \sim {\cal O }
(1),$ then the resulting curvature perturbation after
conversion will be given by \be \z_{conv-end} \approx
\sqrt{\frac{2}{\e_{ek}}}\d s_{ek-end}. \ee We should mention
straight away that the approximations just made will not be
good enough in calculating the non-gaussian corrections to the
linear calculation, but for the present purposes, they will do.
Since the fast-roll parameter $\e_{ek} \sim {\cal O } (10^2)$,
we find that \be \z_{conv-end} \sim \frac{1}{10} \d s_{ek-end}.
\ee

\begin{figure}[t]
\begin{center}
\includegraphics[width=0.75\textwidth]{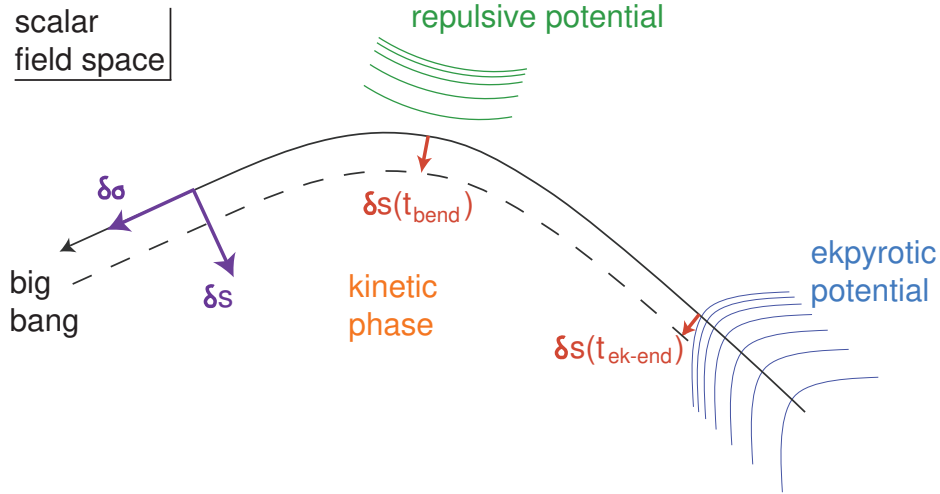}
\caption{\label{figurekineticphase} {\small After the ekpyrotic
phase, the trajectory in scalar field space enters the
kinetic phase and bends - this bending is described by the
existence of an effective repulsive potential (the
potentials are indicated by their contour lines). A
trajectory adjacent to the background evolution can be
characterized by the entropy perturbation
$\d s(t_{ek-end})$ at the end of the ekpyrotic phase,
leading to a corresponding off-set $\d s(t_{bend}),$
or equivalently $\d V(t_{bend}),$ at the time of bending.
}}
\end{center}
\end{figure}

The second way in which a bending of the trajectory can
occur is suggested by the embedding of the cyclic model
in M-theory \cite{Lehners:2007ac}, and applies when the background field trajectory remains straight throughout the ekpyrotic phase. At the end of the
ekpyrotic phase, the potential turns off, and the evolution
becomes dominated by the kinetic energy of the two scalar
fields. This kinetic phase corresponds to the final
approach of the branes in the higher-dimensional picture.
During this approach, there is a generic effect that
occurs, but that cannot be seen in the 4-dimensional
effective theory. The branes that are bound to collide
with each other are of opposite tension. Now, it turns
out that just before the collision, it always happens
that at the location of the negative tension brane, the
internal 6-dimensional manifold tries to shrink to zero
size \cite{Lehners:2006pu}. However, just about any type of matter present on
the negative tension brane will smoothly cause the
internal manifold to grow again \cite{Lehners:2007nb}.
This effect is due to the special properties of gravity
on a negative tension object. When this effect is translated
back into the effective theory that we have been using all
along, the location in field space where the internal
manifold reaches zero size is given by the $\phi_2=0$ line.
This line thus constitutes a boundary to field space. And
the presence of matter on the negative tension brane gives
rise to en effective repulsive potential in the vicinity
of the $\phi_2=0$ line. Hence, during the kinetic phase,
the background trajectory automatically bends, just before
the trajectory shoots off to $-\infty$ where the brane
collison/big bang occurs. What we have just discussed is a
concrete example originating from string theory of how a
bend in the trajectory can occur during the kinetic phase.
However, more generally our results will apply whatever the
microphysical cause of the bending and of the effective
repulsive potential, see Fig. \ref{figurekineticphase} for
an illustration of the general case that we have in mind.
Again, it is quite straightforward to estimate the amplitude
of the curvature perturbation resulting from this process of {\it kinetic conversion}: \\
On large scales, the linearized equation of motion for the
entropy perturbation is given by \be \ddot{\delta s} +
3H\dot{\delta s} + \left(V_{ss} + 3 \dot\th^2 \right) \delta s
= 0,\label{eq-entropy-linearlargescales} \ee where,
incidentally, we have the useful relation
$\dot\th=-V_{s}/\dot\s.$ Then, during the kinetic phase and
away from the repulsive potential, the Einstein equations
immediately yield \be H=\frac{1}{3t}, \qquad
\dot\s=\frac{-\sqrt{2}}{\sqrt{3} t}. \qquad \text{(kinetic
phase)}\label{kineticbackground}\ee Thus, (\ref{eq-entropy-linearlargescales})
simplifies to $\ddot{\delta s} + \dot{\delta s}/t=0$ which
implies that the entropy perturbation grows logarithmically
during the potential-free kinetic phase. We can ignore this
insignificant growth. However, the evolution of the entropy
perturbation during the process of conversion turns out to be
important. We can estimate it by assuming that the trajectory
bends with a constant $\dot\th \sim 1/\Delta t,$ where $\Delta
t$ denotes the duration of the conversion process. We can
further assume that the repulsive potential depends only on
$\phi_2.$ Then $\dot\th, V_{s}, V_{ss}$ can all be related to
$V_{,\phi_2},V_{,\phi_2 \phi_2},$ evaluated during the
conversion, and it is not difficult to show that this leads to
$V_{ss}\approx \dot\phi_1 \dot\th/(t_{bend}\dot\phi_2),$ where
$t_{bend}$ corresponds to the time halfway through the bending
of the trajectory \cite{Lehners:2008my}. For the particular
example where the cyclic model is embedded in
M-theory, we have that $\dot\phi_1=-\sqrt{3}\dot\phi_2,$ and
specializing to this example, we have $V_{ss}\approx
(2-3)\dot\th^2.$ Hence, (\ref{eq-entropy-linearlargescales})
becomes (where we can neglect the term in $\dot{\d s}$) \be
\ddot{\delta s}  + 6 \dot\th^2 \delta s \approx 0, \ee and thus,
during the conversion, the entropy perturbation evolves
sinusoidally \be \d s \approx \cos[\omega(t-t_{conv-beg})]\d
s(t_{ek-end}),\label{entropyduringconversion}\ee where $t_{conv-beg}$ denotes the time at
which the trajectory starts to bend, and $\omega\approx
2.5/\Delta t.$ Now we can immediately evaluate the resulting
linear curvature perturbation by integrating Eq.
(\ref{zetalinear}) to get \bea \z_L &=& \int_{bend}
-\frac{2H}{\dot\s}\dot\th \d s \\ &\approx& \sqrt{\frac{2}{3}}
\frac{\dot\th}{\omega} \sin(\omega \Delta t) \d s(t_{ek-end})
\\ &\approx& \frac{1}{5} \delta s(t_{ek-end}).
\label{zetaconverted} \eea Thus, the amplitude is very similar
in magnitude to the value estimated above for the process of
ekpyrotic conversion.

We are now in a position to calculate the variance of the
generated curvature perturbation, which, on account of (\ref{entropylineargenerated}), is given by \be \langle
\z^2 \rangle \approx \int \frac{d^3 k}{(2\pi)^3}\frac{\e_{ek}
V_{ek-end}}{50 k^{2\n}} = \int \frac{dk}{k} \frac{\e_{ek}
V_{ek-end}}{100\pi^2} k^{n_s-1}.\ee Hence, the amplitude is in
agreement with the current WMAP bounds of
$\Delta_\z^2(0.002Mpc^{-1})=(2.4\pm 0.2)\times 10^{-9}$ \cite{Komatsu:2008hk}, as
long as $|V_{ek-end}|\approx (10^{-2} M_{Pl})^4,$ \ie the
minimum of the potential has to be roughly at the grand unified
scale for models using kinetic conversion
\cite{Lehners:2007ac}. This scale is also the natural scale of
Ho\v{r}ava-Witten theory, and thus it is the scale where one
would expect the potential to turn around. Note that for models
using ekpyrotic conversion, this result implies that the
bending must occur at a specific time, namely when the
potential reaches the grand unified scale. In the latter
models, this may or may not also correspond to the bottom of
the potential.

Finally, we should state an important assumption that we have
been making implicitly up to now: namely, we assumed that the
curvature perturbation passes through the big crunch/big bang
transition essentially unchanged. The reason for doing so is
that the perturbations we are considering are vastly larger
than the horizon size around the time of the crunch, and hence,
due to causality, it seems reasonable to assume that
long-wavelength modes suffer no change - this viewpoint is
discussed in much more detail in \cite{Creminelli:2004jg}. In
new ekpyrotic models, in which the bounce is smooth and
describable entirely within a 4-dimensional effective theory,
this assumption certainly holds true. In the case of a
classically singular bounce, this remains an assumption subject
to possible revision in the future\footnote{In this context, we
can also mention the possibility that no conversion of entropy
to curvature perturbations might occur before the big crunch,
but that this conversion could happen during the phase shortly
following the bang through modulated reheating
\cite{Battefeld:2007st}: if massive matter fields are produced
copiously at the brane collision and dominate the energy
density immediately after the bang, and if, furthermore, these
fields couple to ordinary matter via a function of $\delta s$,
then their decay into ordinary matter will occur at slightly
different times depending on the value of $\delta s.$  In this
way, the ordinary matter perturbations would also inherit the
entropic perturbation spectrum.}.

\section{Higher-Order Perturbations and Predictions for Non-Gaussianity}

\subsection{Definitions and Local Non-Gaussianity}

Now that we have seen in detail how the ekpyrotic phase
generates linear, nearly scale-invariant density perturbations
via the entropic mechanism, we can inquire as to whether the
higher-order corrections might lead to an observable signal. We
will only calculate non-gaussian corrections for perturbations
generated via the entropic mechanism, because, as discussed in
the previous section, this is the only robust and
well-understood mechanism to date that generates ekpyrotic
perturbations in agreement with observations. As we saw
earlier, the linear perturbations are related to observations
of the 2-point correlation function. Similarly, quadratic and
cubic corrections to these perturbations are related to
observations of the 3- and 4-point functions respectively. For
an exactly gaussian probability distribution, all $n$-point
functions for which $n$ is odd vanish, while for $n$ even, the
$n$-point functions are related to the 2-point function. Thus,
the simplest way in which we could detect a departure from
exact gaussianity would be due to the presence of a
non-vanishing 3-point function.

In momentum space, the 3-point function corresponds to a
configuration of 3 momenta, which form a closed triangle due to
momentum conservation. Hence, the 3-point function is specified
not only by its magnitude on different scales, but also by its
magnitude for different shapes of the triangle. Or, turning
this reasoning around, when we make predictions for
non-gaussianity, we must predict both the amplitude and the
shape of the momentum space triangle that we would like to
observe. Let us make all of this more precise now. Earlier, we
defined the power spectrum as the Fourier transform of the
2-point function, \be \langle \z_{\bf k_1}\z_{\bf k_2}\rangle =
(2\pi)^3 \d^3({\bf k}_1+{\bf k}_2)P(k_1).
\label{defpowerspectrum}\ee Similarly, the {\it bispectrum},
which is the Fourier transform of the 3-point function, is
given by \be \langle \z_{\bf k_1}\z_{\bf k_2}\z_{\bf
k_3}\rangle = (2\pi)^3 \d^3({\bf k}_1+{\bf k}_2+{\bf
k}_3)B(k_1,k_2,k_3), \label{defbispectrum}\ee the {\it
trispectrum}, the Fourier transform of the 4-point function,
via \be \langle \z_{\bf k_1}\z_{\bf k_2}\z_{\bf k_3}\z_{\bf
k_4}\rangle_c = (2\pi)^3 \d^3({\bf k}_1+{\bf k}_2+{\bf
k}_3+{\bf k}_4)T(k_1,k_2,k_3,k_4), \ee and so on. The
$\d$-functions result from momentum conservation, while $B$ and
$T$ are shape functions (for a triangle and a quadrangle
respectively). In the last expression, the subscript $c$
indicates that we only need to consider the connected part of
the 4-point function, {\it i.e.} the part that is not captured
by products of 2-point functions.

For non-gaussianity of the so-called {\it local} form, it is
useful to define (in real space) the following expansion of the
curvature perturbation on uniform energy density surfaces \be
\z= \z_L + \frac{3}{5}f_{NL} \z_L^2 + \frac{9}{25}g_{NL}
\z_L^3, \label{zetaexpansion}\ee with $\z_L$ being the linear,
gaussian part of $\z.$ The factors of $3/5$ are a historical
accident; they arose because this type of expansion was first
defined for a different variable.  In momentum space, $B$ is
then given by \be B=\frac{6}{5}f_{NL}[P(k_1)P(k_2) +{\rm 2 \,
permutations}], \label{localbispectrum} \ee as can be verified
straightforwardly by combining Eqs. (\ref{defpowerspectrum}),
(\ref{defbispectrum}) and (\ref{zetaexpansion}). Similarly, the
momentum space 4-point function corresponding to
non-gaussianity of the local form can be expressed as \be
T=\tau_{NL}[P(k_{13})P(k_3)P(k_4) + {\rm 11 \,
perms.}]+\frac{54}{25}g_{NL}[P(k_2)P(k_3)P(k_4)+{\rm 3 \,
perms.}], \ee where $\tau_{NL}$ and $g_{NL}$ parameterize the
two relevant shape functions,  see for example
\cite{Byrnes:2006vq} for more details. For cosmological models
in which the perturbations originate from the fluctuations of a
single field (in our case the entropy field), $\tau_{NL}$ is
directly related to the square of $f_{NL},$ explicitly \be
\tau_{NL} = \frac{36}{25} f_{NL}^2. \label{tauNL}\ee
Concentrating now on the bispectrum, we can see that, since for
a scale-invariant spectrum $P(k)\sim k^{-3},$ we have \bea
B&\sim & f_{NL} (\frac{1}{k_1^3 k_2^3}+\frac{1}{k_2^3
k_3^3}+\frac{1}{k_3^3 k_1^3}) \\ &= & f_{NL} \frac{\Sigma
k_i^3}{\Pi k_i^3}. \eea This is the typical momentum dependence
for local non-gaussianity \cite{Babich:2004gb}, which is also
the relevant one for ekpyrotic models, as we will show shortly.
The signal is largest when one of the momenta is very small -
this automatically requires the other two momenta to be almost
equal, and hence the local form of non-gaussianity corresponds
to having the largest signal generated for {\it squeezed}
triangles in momentum space.

It is instructive to calculate explicitly the tree-level
3-point function for the entropy perturbation generated during
the ekpyrotic phase, \ie before the conversion to curvature
perturbations \footnote{Note that quantum corrections from loop
diagrams will be suppressed by factors of $\hbar.$}. Maldacena
described in \cite{Maldacena:2002vr} how the expectation value
for the 3-point function is given by \be \langle (\d
s)^3\rangle = -i\int^t dt' \langle [(\d
s)^3(t),H_{int}(t')]\rangle, \ee where $H_{int}(t')=V_{sss}(\d
s)^3/3!=-\sqrt{\e}\k_3/(3!t^{'2})$ is the cubic interaction
Hamiltonian. In Fourier space, this can be rewritten as
\cite{Creminelli:2007aq,Koyama:2007if} \bea \langle (\d
s)^3\rangle &=& (2\pi)^3 \d(\Sigma_i {\bf k}_i) \nn \\
&& \times 2{\rm Re}\{ -i\d s_{{\bf k}_1}(t) \d s_{{\bf k}_2}(t)
\d s_{{\bf k}_3}(t) \int_{-\infty}^t
\frac{(-\sqrt{\e}\k_3)}{t^{'2}} \d s_{{\bf k}_1}(t') \d s_{{\bf
k}_2}(t') \d s_{{\bf k}_3}(t') \}. \eea For $\e$ large, the mode
functions are given approximately by ({\it cf.} Eq.
(\ref{modefct})) \be \d s_{{\bf k}}(t) = \frac{1}{\sqrt{2k}}
e^{-ikt}(1-\frac{i}{kt}), \ee so that we get \be \langle (\d
s)^3\rangle = (2\pi)^3 \d(\Sigma_i {\bf k}_i) \frac{\sqrt{\e}
\k_3}{6t^4} \frac{\Sigma k_i^3}{\Pi k_i^3}, \label{entropy3pt}
\ee where we have used the result that \bea && i(1+ik_1
t)(1+ik_2 t)(1+ik_3 t)e^{-iKt}t \int (1-ik_1 t')(1-ik_2
t')(1-ik_3 t')e^{iKt'}(t')^{-5} + c.c.\nn \\ && =
\frac{1}{2}\Sigma_i k_i^3 +\cdots, \eea with $K=k_1+k_2+k_3$
and where the dots indicate terms suppressed by powers of $k_i
t.$ The calculation shows two things: first, the momentum
dependence in (\ref{entropy3pt}) is of the local form, and,
comparing with (\ref{localbispectrum}) and using $P_{\d
s}(k)=1/(2 t^2 k^3)$ from (\ref{spectrumvariance}) and
(\ref{varaincesinglefield}), it corresponds to having an
entropy perturbation of the form \be \d s = \d s_L +
\frac{\sqrt{\e} \k_3}{8} \d s_L^2, \label{entropylocal}\ee
where $\d s_L$ is the linear, gaussian part of the entropy
perturbation $\d s.$ Secondly, the end-result is dominated by
the terms for which $|k_i t| \ll 1$ - in other words, the
dominant contribution to non-gaussianity is generated by the
(classical) evolution on super-horizon scales, and the same
holds true for the 4-point function also. Hence, in determining
the predictions for the non-gaussian curvature perturbation, we
can simply perform the calculation using the higher-order
classical equations of motion on large scales, up to the
required order in perturbation theory.

The strategy for calculating the non-linearity parameters
defined in (\ref{zetaexpansion}) is straightforward: first we
solve the equation of motion for the entropy perturbation up to
third order in perturbation theory. This allows us to integrate
the equation of motion for $\zeta$, at the first three orders
in perturbation theory, and then we obtain the non-linearity
parameters by evaluating  \be f_{NL} = \frac{5}{3}
\frac{\int_{t_{ek-beg}}^{t_{conv-end}}\z^{(2)'}}{(\int_{t_{ek-beg}}^{t_{conv-end}}
\z^{(1)'})^2}, \qquad  g_{NL} = \frac{25}{9}
\frac{\int_{t_{ek-beg}}^{t_{conv-end}}\z^{(3)'}}{(\int_{t_{ek-beg}}^{t_{conv-end}}
\z^{(1)'})^3}, \label{calcstrat}\ee where the integrals are
evaluated from the time that the ekpyrotic phase begins until
the conversion phase has ended and $\zeta$ has evolved to a
constant value. A note on notation: we expand the entropy
perturbation (and similarly the curvature perturbation) as \be
\d s= \d s^{(1)} + \d s^{(2)} + \d s^{(3)} \ee without
factorial factors and where we use $\d s^{(1)}$ and $\d s_L$
interchangeably (in section \ref{sectionlinear}, where we dealt
exclusively with linear perturbations, we sometimes wrote $\d
s$ instead of $\d s^{(1)},$ but hopefully this will not confuse
the reader).

The relevant equations of third order cosmological perturbation
theory with multiple scalar fields was developed in
\cite{Lehners:2009ja}, and we will use the results of that
paper. The derivations of the equations are lengthy, and do not
provide any further insight into the topic of this review.
Hence, we will simply use the equations as we need them.
Interestingly, it turns out that all results regarding the
conversion process can be well approximated by various simple
techniques that we will present further below, both for
ekpyrotic and for kinetic conversion.

\subsection{Generation of Entropy Perturbations}

During the ekpyrotic phase, the equation of motion for the
entropy perturbation, up to third order in perturbation theory,
is given by \cite{Lehners:2009ja} \bea && \ddot{\d s} +
3H\dot{\d
s} +  V_{,ss} \d s + \frac{1}{2}V_{,sss}(\d s)^2 \nn \\
&& +\frac{V_{,\s}}{\dot\s^{3}}(\dot{\d s})^3
+\left(\frac{V_{,\s\s}}{\dot\s^{2}}
+3\frac{V_{,\s}^2}{\dot\s^{4}}+3H\frac{V_{,\s}}{\dot\s^{3}}-2\frac{V_{,ss}}{\dot\s^{2}}\right)
(\dot{\d s})^2 \d s \nn \\ && +\left(-\frac{3}{2\dot\s}
V_{,ss\s}-5\frac{V_{,\s}V_{,ss}}{\dot\s^{3}}
-3H\frac{V_{,ss}}{\dot\s^{2}}\right)\dot{\d s} (\d s)^2
+\left(\frac{1}{6}V_{,ssss}+2\frac{V_{,ss}^2}{\dot\s^{2}}
\right)(\d s)^3 = 0 \,. \label{s3'}\eea Using the following
expressions, valid during the ekpyrotic phase, \bea
&& \dot\s = -\frac{\sqrt{2}}{\sqrt{\e}t} \qquad V= -\frac{1}{\e t^2} \nn \\
&& V_{,\s}= -\frac{\sqrt{2}}{\sqrt{\e}t^2} \qquad
V_{,\s\s}= -\frac{2}{t^2} \qquad V_{,s\s} = 0
\qquad
V_{,ss\s} = -\frac{2\sqrt{2\e}}{t^2} \nn \\
&& V_{,s}= 0 \qquad V_{,ss} = -\frac{2}{t^2} \qquad V_{,sss}=
-\frac{\k_3 \sqrt{\e}}{t^2}    \qquad V_{,ssss}=- \frac{\k_4
\e}{t^2},  \nn \eea it is not difficult to find by iteration
that the entropy perturbation, to leading order in $1/\e$, is
given by \be \d s= \d s_L + \frac{\k_3 \sqrt{\e}}{8}\d s_L^2 +
\e(\frac{\k_4}{60}+\frac{\k_3^2}{80}-\frac{2}{5})\d s_L^3,
\label{entropyInitialCondition}\ee where, just as before, $\d
s_L \propto 1/t.$ Note that the quadratic term agrees exactly
with (\ref{entropylocal}). At each order in perturbation
theory, the non-linear corrections depend on the parameters of
the potential at that order, {\it cf.} the parametrization of
the potential in Eq. (\ref{potentialParameterized}). The above
equation specifies the initial conditions for the start of the
conversion phase.

Both during the ekpyrotic phase and during the conversion
process, when the background trajectory bends, the entropy
perturbation sources the curvature perturbation on large scales
according to \cite{Lehners:2009ja} \bea \dot\z &\approx&
\frac{2H}{\dot\s^{2}} [V_{,s} \d s-\frac{1}{2\dot\s}
V_{,\s}\d s \dot{\d s}+(\frac{1}{2}V_{,ss}+\frac{2}{\dot\s^2}V_{,s}^2)(\d s)^2] \nn \\ && +\frac{2H}{\dot\s^{2}}
[(-\frac{\dot\th}{6\dot\s^{2}} V_{,\s}-
\frac{1}{2\dot\s}V_{,s\s}-\frac{2}{\dot\s^3} V_{,s} V_{,\s})(\d
s)^2 \dot{\d s} + (\frac{1}{6}V_{,sss}+ \frac{2}{\dot\s^2} V_{,s} V_{,ss}+\frac{4}{\dot\s^{4}} V_{,s}^3) (\d s)^3].
\label{zeta3'}
\eea The inelegant, but
sure-fire thing to do now is to simply integrate this equation
numerically during the two phases of generation and conversion,
and to deduce the non-linearity parameters using
(\ref{calcstrat}), as was done in \cite{Lehners:2009ja}.
However, this approach does not give much insight into the
final result. This is why we will present more physically
transparent techniques first, which allow us to estimate the
non-linearity parameters $f_{NL}$ and $g_{NL}$ pretty well, and
subsequently we will compare these estimates with the results
of numerical integration and discuss the predictions.

\subsection{Ekpyrotic Conversion}

We start by analyzing the case where the conversion of entropy
into curvature modes occurs during the ekpyrotic phase. For
this scenario, it was shown by Koyama {\it et al.}
\cite{Koyama:2007if} that the so-called $\d N$ formalism \cite{Starobinsky:1986fxa,Sasaki:1995aw} is
well suited. For ekpyrotic conversion, the calculation is most
easily performed, and the result most easily expressed, in
terms of the potential (\ref{potential2field}), which is why we
are adopting this restricted form here. In working with a
parameterized potential like (\ref{potentialParameterized}),
the bending of the trajectory can be more complicated, in the
sense that there can be multiple turns, and one has to decide
when to stop the evolution. In this case, the results are
strongly cut-off dependent, and without a precisely defined
model specifying the subsequent evolution, it is impossible to
make any generic predictions.

The $\d N$ formalism is particularly well suited to the case of
ekpyrotic conversion, because the background evolution is
simple. In fact, it turns out that by making the approximation
that the bending is instantaneous, it is very easy to find an
approximate formula for the non-linearity parameters at any
chosen order in perturbation theory. The following calculation
was first presented in \cite{Koyama:2007if} for the bispectrum,
and trivially extended to the trispectrum in
\cite{Lehners:2009ja}.

In order to implement the $\d N$ formalism, we have to
calculate the integrated expansion $N = \int H dt$ along the
background trajectory. Initially, the trajectory is close to
the scaling solution (\ref{ScalingSolution}). Then, we assume
that at a fixed field value $\d s_B$ away from the ridge, the
trajectory instantly bends and rolls off along the $\phi_2$
direction. At this point, the trajectory follows the
single-field evolution \be a(t)=(-t)^{2/c_2^2} \qquad \phi_2 =
\frac{2}{c_2} \ln(-t) + {\rm \, constant} \qquad \phi_1 = {\rm
constant}. \label{solutionSingleField}\ee Approximating the
bending as instantaneous, it is easy to evaluate the integrated
expansion, with the result that \be N = -\frac{2}{c_1^2}
\ln|H_B| + {\rm \, constant}, \ee where $H_B$ denotes the
Hubble parameter at the instant that the bending occurs. Note
that all $c_2$-dependence has canceled out of the formula
above. At the end of the conversion process, we are interested
in evaluating the curvature perturbation on a surface of
constant energy density. But, on a surface of constant energy
density, the curvature perturbation is equal to a perturbation
in the integrated expansion \cite{Lyth:2004gb}. If we assume
that the integrated expansion depends on a single variable
$\a$, we can write \be \zeta= \d N = N_{,\a} \d \a +
\frac{1}{2} N_{,\a\a} (\d \a)^2 + \frac{1}{6} N_{,\a\a\a} (\d
\a)^3. \ee In our example, we indeed expect a change in $N$ to
depend solely on a change in the initial value of the entropy
perturbation $\d s.$ Now, from Eq.
(\ref{entropyInitialCondition}), we know that $\d s_L \propto
1/t \propto H,$ and hence we can parameterize different initial
values of the entropy perturbation by writing \be \d s_L = \a
H. \ee Note that since $\d s_L$ is gaussian, so is $\a.$ With
this identification, we have \be \d s= \a H + \frac{\k_3
\sqrt{\e}}{8}(\a H)^2 +
\e(\frac{\k_4}{60}+\frac{\k_3^2}{80}-\frac{2}{5})(\a H)^3, \ee
so that at the fixed value $\d s = \d s_B,$ we have \be \a
\propto \frac{1}{H_B}. \ee Now we can immediately evaluate \be
N_{,\a} = N_{,H_B}\frac{d H_B}{d\a} = \frac{2}{c_1^2 \a},\ee
and similarly \be N_{,\a\a} = -\frac{2}{c_1^2 \a^2} \qquad
N_{,\a\a\a} = \frac{4}{c_1^2 \a^3}. \ee In this way, with very
little work, we can estimate the non-linearity parameters \bea
f_{NL} &=&
\frac{5N_{,\a\a}}{6N_{,\a}^2} = -\frac{5}{12} c_1^2 \\
\tau_{NL} &=& \frac{36}{25}f_{NL}^2 = \frac{1}{4} c_1^4 \\
g_{NL} &=& \frac{25 N_{,\a\a\a}}{54 N_{,\a}^3} =
\frac{25}{108}c_1^4. \eea We are now in a position to compare
these estimates to the numerical results obtained in
\cite{Lehners:2009ja} by solving and integrating the equations
of motion (\ref{s3'}) and (\ref{zeta3'})\footnote{Approximate
analytic solutions to the equations of motion were first
presented in \cite{Buchbinder:2007at}.}. There, the initial
conditions were chosen such that they are given by the scaling
solution (\ref{ScalingSolution}), except for an $0.1$ percent
increase in the initial field velocity $|\dot\phi_2|.$ This causes
the trajectory to eventually roll off in the $\phi_2$
direction, and to quickly approach the single-field solution
(\ref{solutionSingleField}). The results for several values of
$c_1$ and $c_2$ are shown table \ref{Table1}, alongside the
values estimated by the $\d N$ formulae.

\begin{table}
\begin{tabular}{|c|c||c|c|c||c|c|c|}
  \hline
  $c_1$ & $c_2$ & $f_{NL,\delta N}$ & $\tau_{NL,\d N}$ & $g_{NL, \d N}$ & $f_{NL}$ & $\tau_{NL}$ & $g_{NL}$ \\ \hline
  10 & 10 & -41.67  & 2500 & 2315 & -39.95  & 2298 & 2591 \\
  10 & 15 & -41.67  & 2500 & 2315 & -40.45  & 2356 & 2813 \\
  10 & 20 & -41.67  & 2500 & 2315 & -40.62  & 2377 & 3030 \\
  15 & 10 & -93.75  & 12660 & 11720 & -91.01  & 11930 & 13100 \\
  15 & 15 & -93.75  & 12660 & 11720 & -92.11  & 12220 & 13830 \\
  15 & 20 & -93.75  & 12660 & 11720 & -92.49  & 12320 & 14440 \\
  20 & 10 & -166.7  & 40000 & 37040 & -162.5  & 38020 & 41320 \\
  20 & 15 & -166.7  & 40000 & 37040 & -164.4  & 38930 & 43170 \\
  20 & 20 & -166.7  & 40000 & 37040 & -165.1  & 39240 & 44490 \\
  \hline
\end{tabular}
\caption{\small Ekpyrotic conversion: the values of the
non-linearity parameters estimated by the $\d N$ formalism
(using the instantaneous bending approximation) compared to the
numerical results obtained by directly integrating the
equations of motion, for potentials of the form $V=-V_1 e^{-c_1\phi_1}-V_2 e^{-c_2 \phi_2}$.} \label{Table1}
\end{table}

It is immediately apparent that the general trend is accurately
captured by the $\d N$ formulae. However, one may notice that
the agreement is slightly less good at third order than at
second, and also, that the $\d N$ formulae tend to slightly
over-estimate $\tau_{NL}$ and slightly under-estimate $g_{NL}.$
But given the quickness of the $\d N$ calculation and the
complexity of the third order equations of motion, the
agreement is pretty impressive. Of course, the $\d N$ formula
was derived subject to the instantaneous bending approximation.
Without this approximation, we would expect a numerical scheme
that uses the $\d N$ formalism to yield results in close
agreement with the numerical results.

\subsection{Kinetic Conversion}

As discussed in more detail at the end of section
\ref{sectionlinear}, in models motivated by M-theory a bend in
the trajectory happens naturally just before the big crunch,
during the final approach of the two end-of-the-world branes.
This bend takes place after the ekpyrotic phase has come to an
end and while the evolution of the universe is dominated by the
kinetic energies of the scalar fields - see also Fig.
\ref{figurekineticphase}. Again, there exists a simple approach
to estimate the non-gaussianity parameters of the curvature
perturbation generated by this process of kinetic conversion,
and we will review it here. This simplified approach was first
presented in \cite{Lehners:2009qu}, based on previous work in
\cite{Buchbinder:2007at,Lehners:2007wc,Lehners:2008my,Lehners:2009ja}.

This estimating procedure is based on the fact that the physics
of the kinetic phase is really quite simple, and moreover,
except for the fact that its initial conditions involve the
entropy perturbation $\delta s,$ the kinetic phase has no
memory of the details of the ekpyrotic phase. In particular,
only the total $\delta s$ in (\ref{entropyInitialCondition})
matters, and the way we choose to decompose it into linear,
second- and third-order parts is irrelevant at this point. What
is more, since $\d s_L \ll 1,$ the second and third order terms
in (\ref{entropyInitialCondition}) are highly subdominant and
we can approximate the evolution of $\d s$ by that of the
linear term $\d s_L$ throughout the kinetic phase. This
realization is the first ingredient of the calculation.

The second is a compact and very useful expression for the
evolution of the curvature perturbation $\z$ on large scales
and on surfaces of constant energy density
\cite{Lyth:2004gb,Buchbinder:2007at}: \be \dot\z= \frac{2\bar H
\d V}{\dot{\bar \s}^{2} -2 \d V}, \label{Zetadot}\ee where $\d
V \equiv V(t,x^i)-\bar V(t)$ and a bar denotes a background
quantity. This equation is exact in the limit where spatial
gradients can be neglected, and can thus be expanded up to the
desired order in perturbation theory if required. If expanded,
it corresponds precisely to Eq. (\ref{zeta3'})
\cite{Lehners:2009qu}.

The third and last ingredient of the calculation is the
simple relationship between $\d V$ and $\d s$ during the
conversion process. As we saw earlier, during the ekpyrotic
phase, the curvature perturbation picks up a blue spectrum and
is hence completely negligible on large scales. To be precise,
since $\d V \neq 0$ during ekpyrosis, there is already some
conversion of entropy into curvature perturbations occurring at
this stage. However, this contribution is entirely negligible
compared to the subsequent conversion (note that since
$V_{,s}=0$ during ekpyrosis, $\d V$ starts out at subleading
order), and hence we can take $\z(t_{ek-end})\approx 0$.
Moreover, as we saw when we were discussing the linear
perturbations, at the end of the conversion process $\z$ is
still significantly smaller than $\d s,$ being given by \be
\z_L \approx \frac{1}{5} \d s_L. \label{zetalinearrepeat}\ee
Hence, during the conversion process, we can take the potential
to depend only on $\d s.$ And since the repulsive potential is
monotonic and we are interested in small departures $\d s \ll
1$ from the background trajectory, it is intuitively clear that
$\d V$ is directly proportional to $\d s$ during the bending. A
numerical calculation readily confirms this simple
relationship.\footnote{For completeness, we mention that the
curvature perturbation can also be sourced by perturbations in
the comoving energy density \cite{Langlois:2006vv}. However,
the ekpyrotic phase massively suppresses comoving energy
density perturbations on large scales; since the kinetic phase
is relatively short, they do not have time to grow and become
significant, so that we can neglect them in our calculation -
see also \cite{Lehners:2007wc}.}

During the conversion, the effect of the repulsive potential is
to cause the entropy perturbation to behave approximately
sinusoidally, as shown in Eq. (\ref{entropyduringconversion}). As we will confirm below, during the conversion
process $\d V \ll \dot\s^2,$ so that Eq. (\ref{Zetadot})
simplifies further to \be \dot\z \approx  \frac{2\bar
H}{\dot{\bar\s}^{2}} \d V, \label{Zetadotapprox}\ee which, when integrated and upon use of (\ref{kineticbackground}) immediately reproduces Eq. (\ref{zetalinearrepeat}). But, as
argued above, $\d s$ as a whole must behave approximately in
this way during the conversion phase, and subsequently
analogous relationships must hold
at higher orders too:
\be \zeta^{(2)} \approx \frac{1}{5} \frac{\k_3 \sqrt{\e}}{8}
\d s_L^2, \qquad  \zeta^{(3)} \approx \frac{1}{5} (\frac{\k_4}{60}+\frac{\k_3^2}{80}-\frac{2}{5})\e \d s_L^3.
\ee These expressions immediately allow us to calculate the
non-linearity parameters \bea && f_{NL} \equiv
\frac{5}{3}\frac{\z^{(2)}}{\z_L^2} \approx 3\k_3\sqrt{\e}, \label{fNLkineticestimate}\\ && g_{NL} \equiv
\frac{25}{9}\frac{\z^{(3)}}{\z_L^3} \approx 70(\frac{\k_4}{60}+\frac{\k_3^2}{80}-\frac{2}{5})\e. \label{gNLkineticestimate}\eea Thus, without much work at all, we find these simple estimates for the non-linearity parameters.

Before discussing this result, let us briefly pause to verify
the approximation made in obtaining Eq. (\ref{Zetadotapprox}):
during the kinetic phase, we can rewrite (\ref{Zetadot}) as \be
\dot\z = \frac{t \, \d V}{1-3 t^2 \d V}. \ee The approximation
made above consists in writing $\dot\z \approx t\d V$ and this
leads to $\z \approx \frac{1}{2} t_{bend}^2 \d V(t_{bend}).$
But we know that by the end of the conversion process $\z
\approx \frac{1}{5} \d s$ and hence we find that \be
3t_{bend}^2 \d V \approx \d s \ll 1,\ee which shows that the
approximation is self-consistent and confirms the validity of
(\ref{Zetadotapprox}).

The results above indicate that the non-linearity that was
present in the entropy perturbation gets transferred
straightforwardly (\ie only with an overall numerical coefficient) to the non-linearity in the curvature
perturbation, essentially due to the simplicity of the kinetic phase. This calculation therefore leads us to expect
no significant additional constant terms in
$f_{NL}$ or constants and $\k_3$-dependent terms in
$g_{NL}$; we will see shortly that this expectation is indeed borne out.

\begin{figure}[t]
\begin{center}
\includegraphics[width=0.65\textwidth]{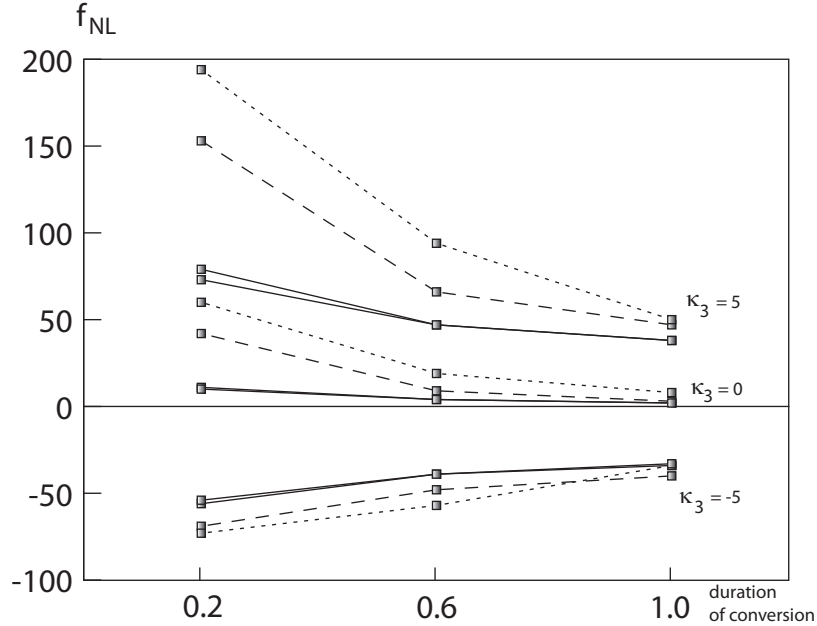}
\caption{\label{FigurefNL} {\small $f_{NL}$ as a function of the duration of conversion and for the values
$\k_3=-5,0,5$ and $\e=36$. In each case, we have plotted the results for four different reflection potentials, with the simplest potentials ($\phi_2^{-2},(\sinh{\phi_2})^{-2}$) indicated by solid lines, while the dashed ($(\sinh{\phi_2})^{-2}+(\sinh{\phi_2})^{-4}$) and dotted ($\phi_2^{-2}+\phi_2^{-6}$) lines give an indication of the range of values that can be expected. As the conversions become smoother, the
predicted range of values narrows, and smooth conversions lead
to a natural range of about $-50 \lesssim f_{NL} \lesssim + 60$ or so.
}}
\end{center}
\end{figure}

\begin{figure}[t]
\begin{center}
\includegraphics[width=0.45\textwidth]{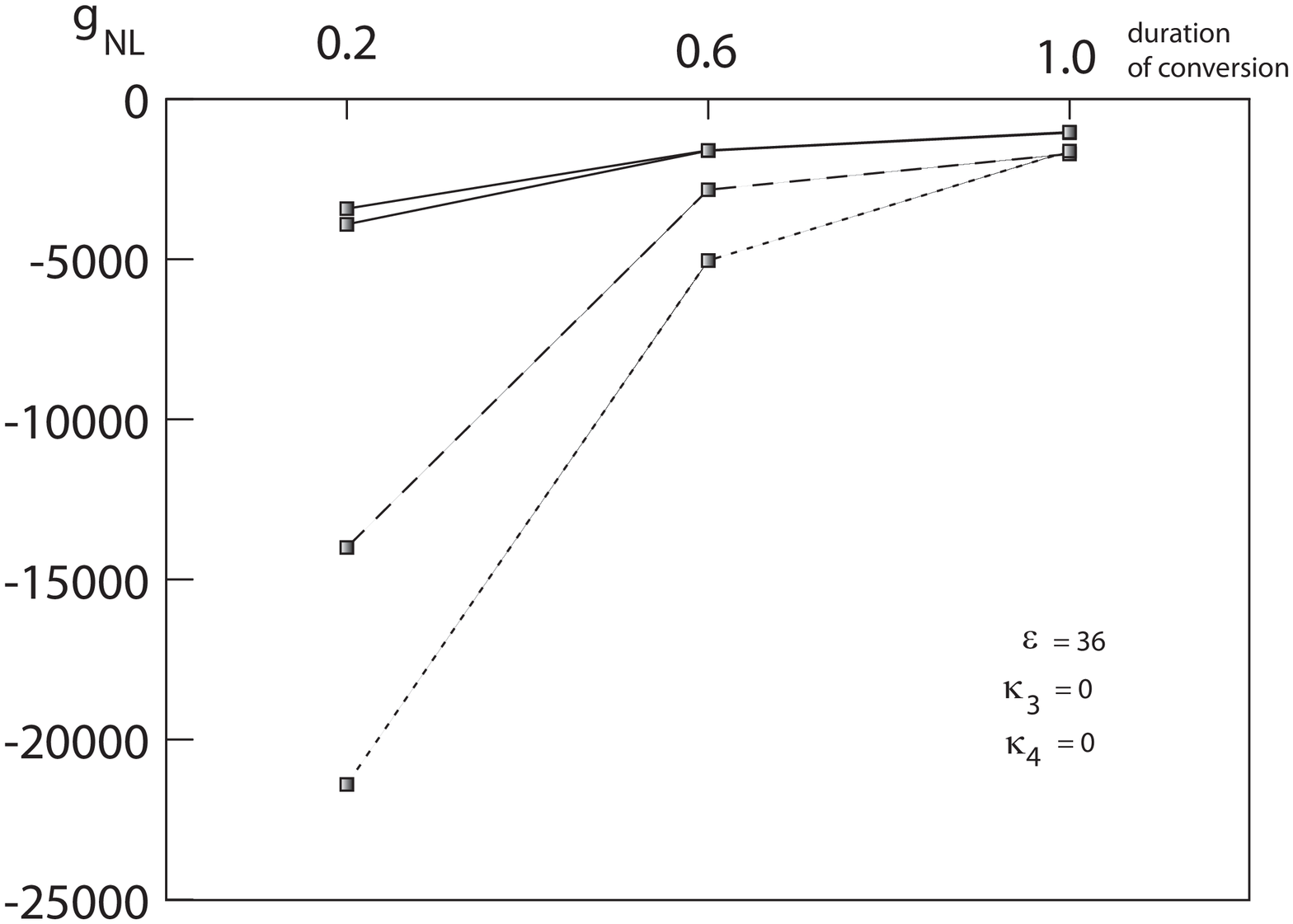} \hfill
\includegraphics[width=0.45\textwidth]{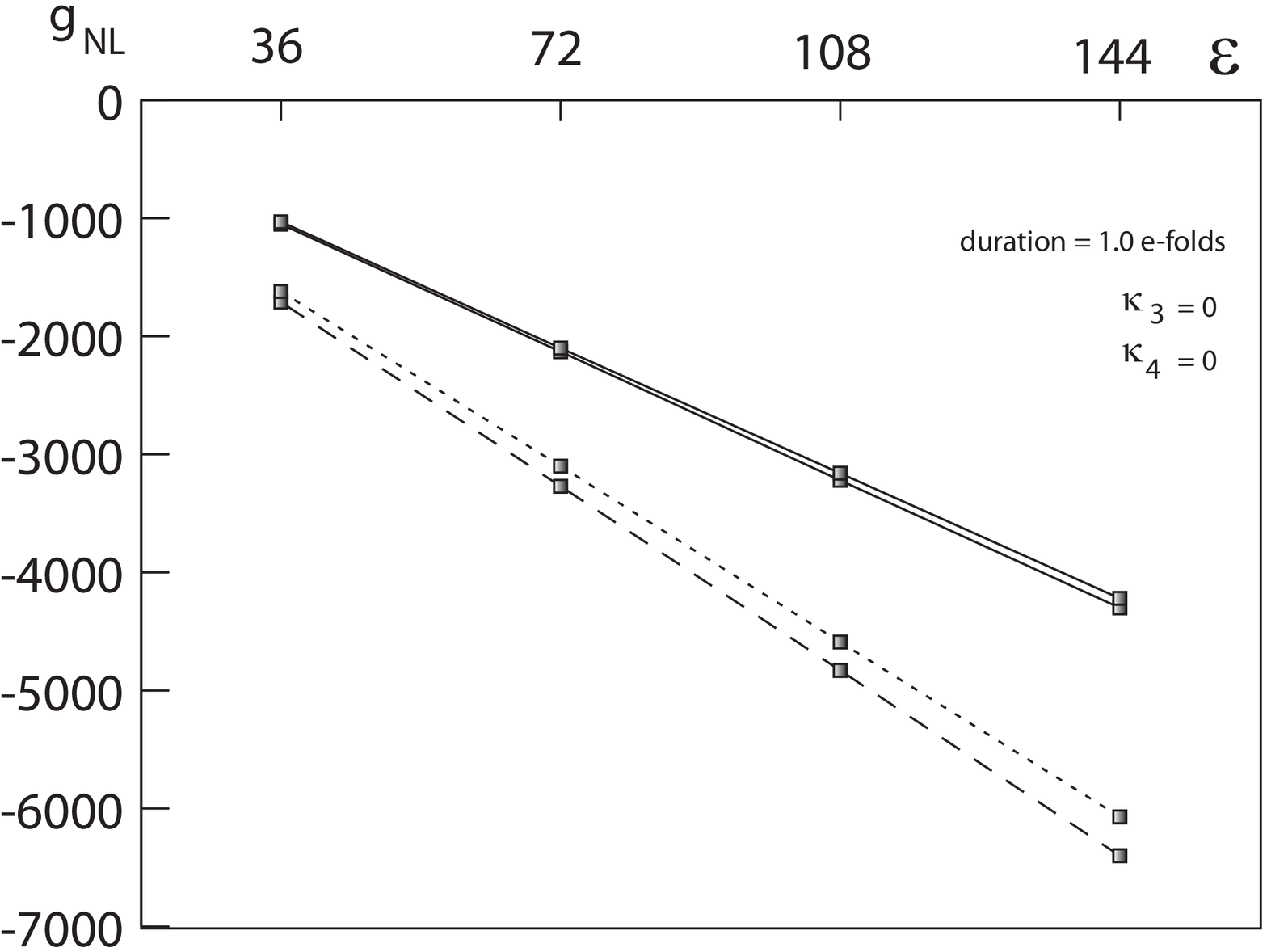}
\caption{\label{Figure1} {\small On the left, $g_{NL}$ is shown as a function of the duration of conversion, with $\k_3=\k_4=0$ and for four different reflection potentials, and with the same line style assignments as in Fig. \ref{FigurefNL}. As the conversions become smoother, the predicted range of values narrows considerably, allowing rather definite predictions. On the right, $g_{NL}$ can be seen to be proportional to $\e$, {\it i.e.} proportional to the equation of state $w_{ek}.$
}}
\end{center}
\end{figure}

\begin{figure}[t]
\begin{center}
\includegraphics[width=0.45\textwidth]{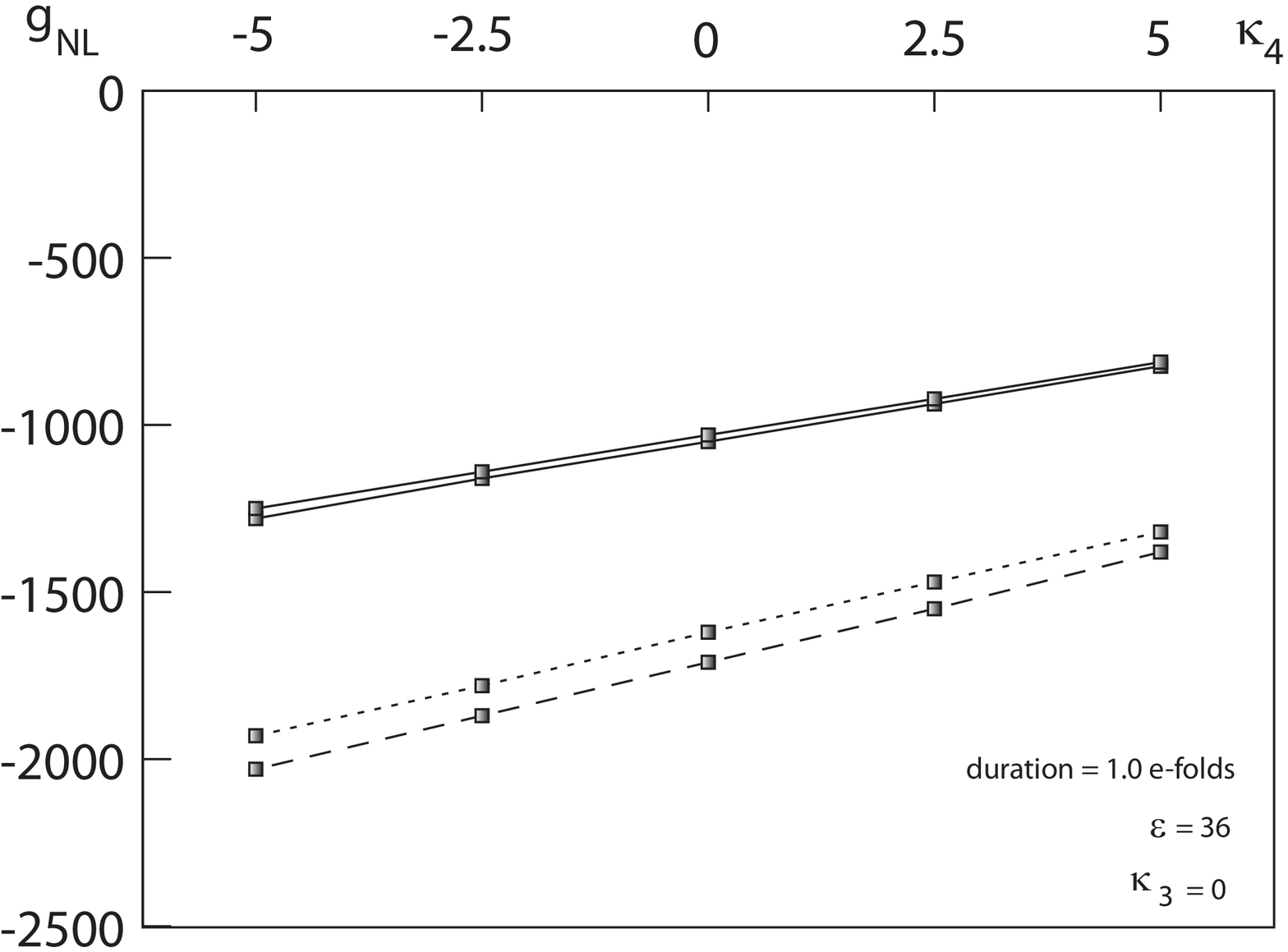} \hfill
\includegraphics[width=0.45\textwidth]{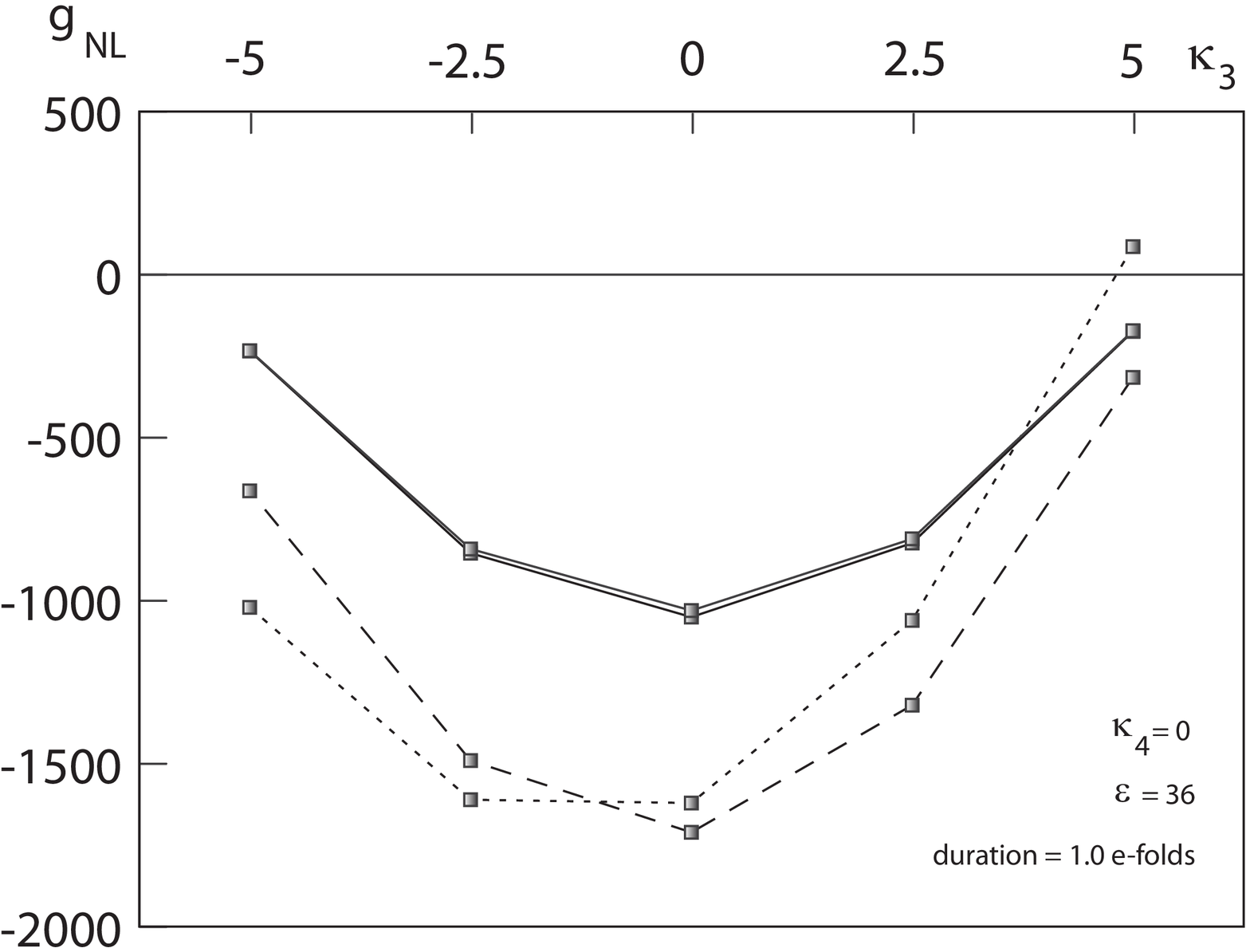}
\caption{\label{Figure3} {\small The left figure indicates that $g_{NL}$ depends linearly in $\k_4,$ the parameter we are using to specify the fourth derivative of the ekpyrotic potential with respect to the entropic direction. Similarly, in the right figure $g_{NL}$ can be seen to depend approximately quadratically on $\k_3,$ the parameter determining the third derivative of the ekpyrotic potential with respect to the entropic direction.
}}
\end{center}
\end{figure}

We can now compare these estimates to the results of brute-force numerical integration \cite{Lehners:2009ja}.
As discussed in section \ref{sectionlinear}, the repulsive
potential can in principle be calculated, given a specific
matter configuration on the negative-tension brane
\cite{Lehners:2007nb}. Here, we
consider four different examples for the repulsive potential \be V_{rep} \propto
\phi_2^{-2},\, \phi_2^{-2}+\phi_2^{-6},\,
(\sinh{\phi_2})^{-2},\,
(\sinh{\phi_2})^{-2}+(\sinh{\phi_2})^{-4},
\label{potentialRepulsive}\ee with the overall magnitude
adjusted in order to obtain various values for the duration of
the conversion (see below). These potential forms should give
an indication of the range of values that one can expect the
non-linearity parameters to take. Note that, without loss of generality, we take the
boundary to be located at $\phi_2=0,$ and we only consider
conversions during which $\dot\th>0$ -- other cases can be related
to these by an appropriate change of coordinates.

An important parameter turns out to be the duration of the
conversion \cite{Lehners:2007wc,Lehners:2009ja}, measured by the number of Hubble times during which most, say 90 percent, of
the conversion takes place, {\it i.e.} the duration is measured by the number of e-folds by which the scale factor shrinks during conversion. Conversions lasting less than about $0.2$ Hubble times
correspond to what we call sharp conversions, while those that last on the order of $1$ Hubble time correspond to smooth conversions. It turns out that the estimating procedure presented above works best for the case of smooth conversions. For
$f_{NL}$, the range of predicted values is considerably
narrower as the conversion becomes smoother
\cite{Lehners:2007wc,Lehners:2008my}, as illustrated in Fig.
\ref{FigurefNL}. Moreover,
$f_{NL}$ can be well fitted by \be f_{NL} = \frac{3}{2} \k_3 \sqrt{\e} + 5,\label{fNLnumerical}\ee in good agreement with the simple estimate (\ref{fNLkineticestimate}).

We will show the results for $g_{NL}$ slightly more explicitly; they are illustrated in Figs.
\ref{Figure1} and \ref{Figure3}. In each case, we have plotted
the results obtained for the four repulsive potentials
(\ref{potentialRepulsive}). The left panel of Fig. \ref{Figure1} shows that, even
more so than for $f_{NL},$ the range of
predicted values for $g_{NL}$ narrows drastically as the
conversion process becomes smoother. In fact, for sharp
conversions, typical values are very large in magnitude, and we
expect these to be observationally disfavored shortly, if they
are not ruled out already. Thus, phenomenologically speaking, it
is much more interesting to focus on smooth
conversions. The right panel in Fig. \ref{Figure1}
shows that $g_{NL}$ is proportional to $\e$, while Fig. \ref{Figure3} indicates that
$g_{NL}$ depends linearly on $\k_4$ and
approximately quadratically on $\k_3$. In fact, all numerical results indicate
that $g_{NL}$ scales with $\e,\k_3,\k_4$ exactly as the
third-order coefficient in the expression
(\ref{entropyInitialCondition}) for the entropy perturbation
during the ekpyrotic phase, and the data can be fitted by the approximate formula \be g_{NL} \approx 100 \,
(\frac{\k_4}{60}+\frac{\k_3^2}{80}-\frac{2}{5}) \, \e,
\label{gNLnumerical}\ee again in good agreement with the estimate (\ref{gNLkineticestimate}).

\section{Discussion of the Results}

In this review, we have focussed on the predictions for non-gaussianity generated via the entropic mechanism in ekpyrotic models and their cyclic extensions. The reason for concentrating on the entropic mechanism is that this is currently the most robust, best-motivated and best-understood mechanism for generating nearly scale-invariant curvature perturbations during a contracting phase of the universe. As indicated by its name, the entropic mechanism achieves this by first generating nearly scale-invariant entropy perturbations, which are subsequently converted into curvature perturbations. There are essentially two distinct possibilities for when this conversion can happen: either directly during the ekpyrotic phase ({\it ekpyrotic conversion}) or during the subsequent phase of scalar field kinetic energy domination right before the big crunch/big bang transition ({\it kinetic conversion}). In both cases, the curvature perturbations pick up non-gaussian corrections of the local form, although the magnitude of the second and third order non-linearity parameters $f_{NL}$ and $g_{NL}$ differ substantially for the two modes of conversion. For convenience, we will repeat the predictions here:\footnote{We should caution the reader that what we are presenting here are the predictions from the ekpyrotic phase alone, \ie we have assumed that the primordial density perturbations generated by the ekpyrotic phase have not been modified by the dynamics of the big crunch/big bang transition, nor that there are additional effects of relevance during the first stages of the subsequent expanding phase of the universe (it is conceivable, for example, that models might be constructed which utilize the ekpyrotic phase, but where there are additional fields that are relevant during the expanding phase and which would contribute to the curvature perturbation - in this case, the final predictions will evidently be model-dependent).}
\bea f_{NL} &=& -\frac{5}{12} c_1^2 \qquad \qquad \qquad {\text{Ekpyrotic Conversion}}\\
g_{NL} &=& \frac{25}{108}c_1^4,\eea
\bea f_{NL} &=& \frac{3}{2}\k_3\, \sqrt{\e} + 5 \qquad \qquad {\text{Kinetic Conversion}}\label{fNLkinetic}\\
g_{NL} &=& (\frac{5}{3}\k_4+\frac{5}{4}\k_3^2-40)\, \e.\label{gNLkinetic}\eea
For ekpyrotic conversion, the results are presented in terms of the parameters of the ekpyrotic potential as expressed in (\ref{potential2field}). The first thing to note is that the signs are unambiguously fixed: $f_{NL}$ is predicted to be always negative, while $g_{NL}$ is always positive. In order for the power spectrum of the perturbations to be in agreement with observations, one needs $c_1\gtrsim 10,$ which implies $f_{NL} \lesssim - 40$ and $g_{NL} \gtrsim 2500.$ The current observational bounds are that $f_{NL} = 38 \pm 21,$ where the errors are quoted at $1\s$ \cite{Smith:2009jr}, while currently no strong constraints exist yet on $g_{NL}.$ These values put the predicted values for $f_{NL}$ for ekpyrotic conversion at $4\s$ or more from the central value, and thus this type of conversion is on the verge of being ruled out by observations. What this means for model-building is that models that have made use of the ekpyrotic conversion mechanism, such as new ekpyrotic models \cite{Buchbinder:2007ad,Creminelli:2007aq}, might have to be modified in a way such that the conversion occurs only once the equation of state becomes small. This is not unnatural, as the ekpyrotic phase must come to an end anyhow before the bounce phase begins, and if the conversion occurs after the end of the ekpyrotic phase, the predictions will be closer to those predicted by the kinetic conversion mechanism, which we turn our attention to now.

For kinetic conversion, the results (\ref{fNLkinetic}) and (\ref{gNLkinetic}) are presented in terms of the parameters of the ekpyrotic potential as written in (\ref{potentialParameterized}). The parameters $\k_3,\k_4$ are expected to be of ${\cal O}(1).$ The fast-roll parameter $\e$ is typically of ${\cal O}(10^2),$ with a lower bound of about $50$ in order for the power spectrum to be in agreement with observations. Thus, the fitting formulae predict the bispectrum parameter $f_{NL}$ to be of order a few tens, with the sign being typically determined by the sign of $\k_3.$ This is in good agreement with current bounds for very natural values of the parameters, such as $\e=100$ and $-1\lesssim \k_3 \lesssim 5,$ for example. The associated prediction for the trispectrum $g_{NL}$ is rather interesting, owing to the fact that the $\k_3,\k_4$-independent contribution is quite large. It sets the ``central'' value of the prediction at $-40 \, \e$ and thus tends to make $g_{NL}$ of order a few thousand and {\it negative} in sign. Even with largish values of $\k_3 \sim 5$ and $\k_4 \sim 5,$ the prediction is still negative $g_{NL} \sim -\e,$ and hence the negative sign of $g_{NL}$ is a rather robust prediction of the kinetic conversion mechanism. For completeness, we note that the second trispectrum parameter $\tau_{NL}$ is always given in terms of $f_{NL}$ according to (\ref{tauNL}), and thus it is predicted to be positive and of order a few hundred. Observational limits on $\t_{NL}$ will thus provide a consistency check of the predictions, and will give an indication whether or not the fluctuations of a single field (in this case the entropy field) were at the origin of the final spectrum of curvature perturbations.

What are the implications of these predictions? Assuming that the observational error bars quoted above will shrink in the near future and that the ekpyrotic conversion mechanism will be ruled out, we will focus here on the predictions of the kinetic conversion mechanism. The most interesting feature is that the natural ranges of the non-linearity parameters are at a level that will be observable by near-future experiments, perhaps already with the PLANCK satellite. Hence, in a few years, we will know whether or not ekpyrotic models, in combination with the entropic mechanism, will be viable or even preferred by the data. In this vein, it is useful to briefly contrast the predictions discussed here with those of inflationary models\footnote{Here, we only contrast the predictions for non-gaussianity of the local form. Non-canonical kinetic terms additionally generate non-gaussianities with different shapes in momentum space, such as equilateral triangles in the case of the bispectrum. This is well understood for inflation, but has not been much explored yet for ekpyrotic models - though see \cite{Khoury:2008wj} for a related study.}: for simple, single-field inflationary models the predicted values for all non-linearity parameters are very small, of order $1$ or smaller \cite{Maldacena:2002vr,Seery:2006js}. Even though these values lie within the predicted ranges of the kinetic conversion mechanism, it is clear that, since these values arise naturally for single-field inflationary models whereas obtaining the same small values in ekpyrosis would require a very finely tuned potential, in case of a non-detection of non-gaussianity the simple inflationary models will be strongly preferred. Conversely, since multi-field inflationary models can give rise to just about any values of the non-linearity parameters \cite{Bartolo:2004if}, in case of a detection of non-gaussianity compatible with the values predicted by the kinetic conversion mechanism, those latter models will be preferred by the data. At that point it will be necessary to include also the observational evidence for or against primordial scale-invariant gravitational waves. Such gravity waves would strongly favor inflationary models, whereas their absence is compatible with ekpyrotic models. The absence of primordial scale-invariant gravitational waves, combined with local non-gaussianity parameters in the ranges predicted by (\ref{fNLkinetic}) and (\ref{gNLkinetic}), would be nothing short of spectacular, as they would point towards the existence of a contracting phase prior to the big bang, and open up the possibility of a multiverse in time, with correspondingly vast timescales unlike anything we are used to in cosmology right now.

\begin{acknowledgements}

I would like to thank S\'{e}bastien Renaux-Petel and Paul Steinhardt for very enjoyable collaborations on the subject of this review. I would also like to thank Justin Khoury, Kazuya Koyama, Burt Ovrut, Neil Turok and David Wands for stimulating and informative discussions.

\end{acknowledgements}


\bibliography{NGReviewLehners}

\begin{thebibliography}{60}
\expandafter\ifx\csname natexlab\endcsname\relax\def\natexlab#1{#1}\fi
\expandafter\ifx\csname bibnamefont\endcsname\relax
  \def\bibnamefont#1{#1}\fi
\expandafter\ifx\csname bibfnamefont\endcsname\relax
  \def\bibfnamefont#1{#1}\fi
\expandafter\ifx\csname citenamefont\endcsname\relax
  \def\citenamefont#1{#1}\fi
\expandafter\ifx\csname url\endcsname\relax
  \def\url#1{\texttt{#1}}\fi
\expandafter\ifx\csname urlprefix\endcsname\relax\def\urlprefix{URL }\fi
\providecommand{\bibinfo}[2]{#2}
\providecommand{\eprint}[2][]{\url{#2}}

\bibitem[{\citenamefont{Baumann}(2009)}]{Baumann:2009ds}
\bibinfo{author}{\bibfnamefont{D.}~\bibnamefont{Baumann}}
  (\bibinfo{year}{2009}), \eprint{0907.5424}.

\bibitem[{\citenamefont{Brustein and Steinhardt}(1993)}]{Brustein:1992nk}
\bibinfo{author}{\bibfnamefont{R.}~\bibnamefont{Brustein}} \bibnamefont{and}
  \bibinfo{author}{\bibfnamefont{P.~J.} \bibnamefont{Steinhardt}},
  \bibinfo{journal}{Phys. Lett.} \textbf{\bibinfo{volume}{B302}},
  \bibinfo{pages}{196} (\bibinfo{year}{1993}), \eprint{hep-th/9212049}.

\bibitem[{\citenamefont{Borde et~al.}(2003)\citenamefont{Borde, Guth, and
  Vilenkin}}]{Borde:2001nh}
\bibinfo{author}{\bibfnamefont{A.}~\bibnamefont{Borde}},
  \bibinfo{author}{\bibfnamefont{A.~H.} \bibnamefont{Guth}}, \bibnamefont{and}
  \bibinfo{author}{\bibfnamefont{A.}~\bibnamefont{Vilenkin}},
  \bibinfo{journal}{Phys. Rev. Lett.} \textbf{\bibinfo{volume}{90}},
  \bibinfo{pages}{151301} (\bibinfo{year}{2003}), \eprint{gr-qc/0110012}.

\bibitem[{\citenamefont{Guth}(2007)}]{Guth:2007ng}
\bibinfo{author}{\bibfnamefont{A.~H.} \bibnamefont{Guth}}, \bibinfo{journal}{J.
  Phys.} \textbf{\bibinfo{volume}{A40}}, \bibinfo{pages}{6811}
  (\bibinfo{year}{2007}), \eprint{hep-th/0702178}.

\bibitem[{\citenamefont{Lehners}(2008)}]{Lehners:2008vx}
\bibinfo{author}{\bibfnamefont{J.-L.} \bibnamefont{Lehners}},
  \bibinfo{journal}{Phys. Rept.} \textbf{\bibinfo{volume}{465}},
  \bibinfo{pages}{223} (\bibinfo{year}{2008}), \eprint{0806.1245}.

\bibitem[{\citenamefont{Seery and Lidsey}(2005)}]{Seery:2005gb}
\bibinfo{author}{\bibfnamefont{D.}~\bibnamefont{Seery}} \bibnamefont{and}
  \bibinfo{author}{\bibfnamefont{J.~E.} \bibnamefont{Lidsey}},
  \bibinfo{journal}{JCAP} \textbf{\bibinfo{volume}{0509}}, \bibinfo{pages}{011}
  (\bibinfo{year}{2005}), \eprint{astro-ph/0506056}.

\bibitem[{\citenamefont{Khoury et~al.}(2001)\citenamefont{Khoury, Ovrut,
  Steinhardt, and Turok}}]{Khoury:2001wf}
\bibinfo{author}{\bibfnamefont{J.}~\bibnamefont{Khoury}},
  \bibinfo{author}{\bibfnamefont{B.~A.} \bibnamefont{Ovrut}},
  \bibinfo{author}{\bibfnamefont{P.~J.} \bibnamefont{Steinhardt}},
  \bibnamefont{and} \bibinfo{author}{\bibfnamefont{N.}~\bibnamefont{Turok}},
  \bibinfo{journal}{Phys. Rev.} \textbf{\bibinfo{volume}{D64}},
  \bibinfo{pages}{123522} (\bibinfo{year}{2001}), \eprint{hep-th/0103239}.

\bibitem[{\citenamefont{Erickson et~al.}(2007)\citenamefont{Erickson, Gratton,
  Steinhardt, and Turok}}]{Erickson:2006wc}
\bibinfo{author}{\bibfnamefont{J.~K.} \bibnamefont{Erickson}},
  \bibinfo{author}{\bibfnamefont{S.}~\bibnamefont{Gratton}},
  \bibinfo{author}{\bibfnamefont{P.~J.} \bibnamefont{Steinhardt}},
  \bibnamefont{and} \bibinfo{author}{\bibfnamefont{N.}~\bibnamefont{Turok}},
  \bibinfo{journal}{Phys. Rev.} \textbf{\bibinfo{volume}{D75}},
  \bibinfo{pages}{123507} (\bibinfo{year}{2007}), \eprint{hep-th/0607164}.

\bibitem[{\citenamefont{Komatsu et~al.}(2009)}]{Komatsu:2008hk}
\bibinfo{author}{\bibfnamefont{E.}~\bibnamefont{Komatsu}} \bibnamefont{et~al.}
  (\bibinfo{collaboration}{WMAP}), \bibinfo{journal}{Astrophys. J. Suppl.}
  \textbf{\bibinfo{volume}{180}}, \bibinfo{pages}{330} (\bibinfo{year}{2009}),
  \eprint{0803.0547}.

\bibitem[{\citenamefont{Buchbinder
  et~al.}(2007{\natexlab{a}})\citenamefont{Buchbinder, Khoury, and
  Ovrut}}]{Buchbinder:2007ad}
\bibinfo{author}{\bibfnamefont{E.~I.} \bibnamefont{Buchbinder}},
  \bibinfo{author}{\bibfnamefont{J.}~\bibnamefont{Khoury}}, \bibnamefont{and}
  \bibinfo{author}{\bibfnamefont{B.~A.} \bibnamefont{Ovrut}},
  \bibinfo{journal}{Phys. Rev.} \textbf{\bibinfo{volume}{D76}},
  \bibinfo{pages}{123503} (\bibinfo{year}{2007}{\natexlab{a}}),
  \eprint{hep-th/0702154}.

\bibitem[{\citenamefont{Buchbinder
  et~al.}(2007{\natexlab{b}})\citenamefont{Buchbinder, Khoury, and
  Ovrut}}]{Buchbinder:2007tw}
\bibinfo{author}{\bibfnamefont{E.~I.} \bibnamefont{Buchbinder}},
  \bibinfo{author}{\bibfnamefont{J.}~\bibnamefont{Khoury}}, \bibnamefont{and}
  \bibinfo{author}{\bibfnamefont{B.~A.} \bibnamefont{Ovrut}},
  \bibinfo{journal}{JHEP} \textbf{\bibinfo{volume}{11}}, \bibinfo{pages}{076}
  (\bibinfo{year}{2007}{\natexlab{b}}), \eprint{arXiv:0706.3903 [hep-th]}.

\bibitem[{\citenamefont{Creminelli and Senatore}(2007)}]{Creminelli:2007aq}
\bibinfo{author}{\bibfnamefont{P.}~\bibnamefont{Creminelli}} \bibnamefont{and}
  \bibinfo{author}{\bibfnamefont{L.}~\bibnamefont{Senatore}},
  \bibinfo{journal}{JCAP} \textbf{\bibinfo{volume}{0711}}, \bibinfo{pages}{010}
  (\bibinfo{year}{2007}), \eprint{hep-th/0702165}.

\bibitem[{\citenamefont{Arkani-Hamed et~al.}(2004)\citenamefont{Arkani-Hamed,
  Cheng, Luty, and Mukohyama}}]{ArkaniHamed:2003uy}
\bibinfo{author}{\bibfnamefont{N.}~\bibnamefont{Arkani-Hamed}},
  \bibinfo{author}{\bibfnamefont{H.-C.} \bibnamefont{Cheng}},
  \bibinfo{author}{\bibfnamefont{M.~A.} \bibnamefont{Luty}}, \bibnamefont{and}
  \bibinfo{author}{\bibfnamefont{S.}~\bibnamefont{Mukohyama}},
  \bibinfo{journal}{JHEP} \textbf{\bibinfo{volume}{05}}, \bibinfo{pages}{074}
  (\bibinfo{year}{2004}), \eprint{hep-th/0312099}.

\bibitem[{\citenamefont{Adams et~al.}(2006)\citenamefont{Adams, Arkani-Hamed,
  Dubovsky, Nicolis, and Rattazzi}}]{Adams:2006sv}
\bibinfo{author}{\bibfnamefont{A.}~\bibnamefont{Adams}},
  \bibinfo{author}{\bibfnamefont{N.}~\bibnamefont{Arkani-Hamed}},
  \bibinfo{author}{\bibfnamefont{S.}~\bibnamefont{Dubovsky}},
  \bibinfo{author}{\bibfnamefont{A.}~\bibnamefont{Nicolis}}, \bibnamefont{and}
  \bibinfo{author}{\bibfnamefont{R.}~\bibnamefont{Rattazzi}},
  \bibinfo{journal}{JHEP} \textbf{\bibinfo{volume}{10}}, \bibinfo{pages}{014}
  (\bibinfo{year}{2006}), \eprint{hep-th/0602178}.

\bibitem[{\citenamefont{Kallosh et~al.}(2008)\citenamefont{Kallosh, Kang,
  Linde, and Mukhanov}}]{Kallosh:2007ad}
\bibinfo{author}{\bibfnamefont{R.}~\bibnamefont{Kallosh}},
  \bibinfo{author}{\bibfnamefont{J.~U.} \bibnamefont{Kang}},
  \bibinfo{author}{\bibfnamefont{A.~D.} \bibnamefont{Linde}}, \bibnamefont{and}
  \bibinfo{author}{\bibfnamefont{V.}~\bibnamefont{Mukhanov}},
  \bibinfo{journal}{JCAP} \textbf{\bibinfo{volume}{0804}}, \bibinfo{pages}{018}
  (\bibinfo{year}{2008}), \eprint{0712.2040}.

\bibitem[{\citenamefont{Steinhardt and Turok}(2001)}]{Steinhardt:2001vw}
\bibinfo{author}{\bibfnamefont{P.~J.} \bibnamefont{Steinhardt}}
  \bibnamefont{and} \bibinfo{author}{\bibfnamefont{N.}~\bibnamefont{Turok}}
  (\bibinfo{year}{2001}), \eprint{hep-th/0111030}.

\bibitem[{\citenamefont{Steinhardt and Turok}(2002)}]{Steinhardt:2001st}
\bibinfo{author}{\bibfnamefont{P.~J.} \bibnamefont{Steinhardt}}
  \bibnamefont{and} \bibinfo{author}{\bibfnamefont{N.}~\bibnamefont{Turok}},
  \bibinfo{journal}{Phys. Rev.} \textbf{\bibinfo{volume}{D65}},
  \bibinfo{pages}{126003} (\bibinfo{year}{2002}), \eprint{hep-th/0111098}.

\bibitem[{\citenamefont{Lukas et~al.}(1999)\citenamefont{Lukas, Ovrut, Stelle,
  and Waldram}}]{Lukas:1998yy}
\bibinfo{author}{\bibfnamefont{A.}~\bibnamefont{Lukas}},
  \bibinfo{author}{\bibfnamefont{B.~A.} \bibnamefont{Ovrut}},
  \bibinfo{author}{\bibfnamefont{K.~S.} \bibnamefont{Stelle}},
  \bibnamefont{and} \bibinfo{author}{\bibfnamefont{D.}~\bibnamefont{Waldram}},
  \bibinfo{journal}{Phys. Rev.} \textbf{\bibinfo{volume}{D59}},
  \bibinfo{pages}{086001} (\bibinfo{year}{1999}), \eprint{hep-th/9803235}.

\bibitem[{\citenamefont{Horava and Witten}(1996)}]{Horava:1995qa}
\bibinfo{author}{\bibfnamefont{P.}~\bibnamefont{Horava}} \bibnamefont{and}
  \bibinfo{author}{\bibfnamefont{E.}~\bibnamefont{Witten}},
  \bibinfo{journal}{Nucl. Phys.} \textbf{\bibinfo{volume}{B460}},
  \bibinfo{pages}{506} (\bibinfo{year}{1996}), \eprint{hep-th/9510209}.

\bibitem[{\citenamefont{Turok et~al.}(2004)\citenamefont{Turok, Perry, and
  Steinhardt}}]{Turok:2004gb}
\bibinfo{author}{\bibfnamefont{N.}~\bibnamefont{Turok}},
  \bibinfo{author}{\bibfnamefont{M.}~\bibnamefont{Perry}}, \bibnamefont{and}
  \bibinfo{author}{\bibfnamefont{P.~J.} \bibnamefont{Steinhardt}},
  \bibinfo{journal}{Phys. Rev.} \textbf{\bibinfo{volume}{D70}},
  \bibinfo{pages}{106004} (\bibinfo{year}{2004}), \eprint{hep-th/0408083}.

\bibitem[{\citenamefont{Lehners
  et~al.}(2007{\natexlab{a}})\citenamefont{Lehners, McFadden, and
  Turok}}]{Lehners:2006pu}
\bibinfo{author}{\bibfnamefont{J.-L.} \bibnamefont{Lehners}},
  \bibinfo{author}{\bibfnamefont{P.}~\bibnamefont{McFadden}}, \bibnamefont{and}
  \bibinfo{author}{\bibfnamefont{N.}~\bibnamefont{Turok}},
  \bibinfo{journal}{Phys. Rev.} \textbf{\bibinfo{volume}{D75}},
  \bibinfo{pages}{103510} (\bibinfo{year}{2007}{\natexlab{a}}),
  \eprint{hep-th/0611259}.

\bibitem[{\citenamefont{Lehners
  et~al.}(2007{\natexlab{b}})\citenamefont{Lehners, McFadden, and
  Turok}}]{Lehners:2006ir}
\bibinfo{author}{\bibfnamefont{J.-L.} \bibnamefont{Lehners}},
  \bibinfo{author}{\bibfnamefont{P.}~\bibnamefont{McFadden}}, \bibnamefont{and}
  \bibinfo{author}{\bibfnamefont{N.}~\bibnamefont{Turok}},
  \bibinfo{journal}{Phys. Rev.} \textbf{\bibinfo{volume}{D76}},
  \bibinfo{pages}{023501} (\bibinfo{year}{2007}{\natexlab{b}}),
  \eprint{hep-th/0612026}.

\bibitem[{\citenamefont{Koyama and Wands}(2007)}]{Koyama:2007mg}
\bibinfo{author}{\bibfnamefont{K.}~\bibnamefont{Koyama}} \bibnamefont{and}
  \bibinfo{author}{\bibfnamefont{D.}~\bibnamefont{Wands}},
  \bibinfo{journal}{JCAP} \textbf{\bibinfo{volume}{0704}}, \bibinfo{pages}{008}
  (\bibinfo{year}{2007}), \eprint{hep-th/0703040}.

\bibitem[{\citenamefont{Koyama et~al.}(2007{\natexlab{a}})\citenamefont{Koyama,
  Mizuno, and Wands}}]{Koyama:2007ag}
\bibinfo{author}{\bibfnamefont{K.}~\bibnamefont{Koyama}},
  \bibinfo{author}{\bibfnamefont{S.}~\bibnamefont{Mizuno}}, \bibnamefont{and}
  \bibinfo{author}{\bibfnamefont{D.}~\bibnamefont{Wands}},
  \bibinfo{journal}{Class. Quant. Grav.} \textbf{\bibinfo{volume}{24}},
  \bibinfo{pages}{3919} (\bibinfo{year}{2007}{\natexlab{a}}),
  \eprint{0704.1152}.

\bibitem[{\citenamefont{Gordon et~al.}(2001)\citenamefont{Gordon, Wands,
  Bassett, and Maartens}}]{Gordon:2000hv}
\bibinfo{author}{\bibfnamefont{C.}~\bibnamefont{Gordon}},
  \bibinfo{author}{\bibfnamefont{D.}~\bibnamefont{Wands}},
  \bibinfo{author}{\bibfnamefont{B.~A.} \bibnamefont{Bassett}},
  \bibnamefont{and} \bibinfo{author}{\bibfnamefont{R.}~\bibnamefont{Maartens}},
  \bibinfo{journal}{Phys. Rev.} \textbf{\bibinfo{volume}{D63}},
  \bibinfo{pages}{023506} (\bibinfo{year}{2001}), \eprint{astro-ph/0009131}.

\bibitem[{\citenamefont{Lehners
  et~al.}(2007{\natexlab{c}})\citenamefont{Lehners, McFadden, Turok, and
  Steinhardt}}]{Lehners:2007ac}
\bibinfo{author}{\bibfnamefont{J.-L.} \bibnamefont{Lehners}},
  \bibinfo{author}{\bibfnamefont{P.}~\bibnamefont{McFadden}},
  \bibinfo{author}{\bibfnamefont{N.}~\bibnamefont{Turok}}, \bibnamefont{and}
  \bibinfo{author}{\bibfnamefont{P.~J.} \bibnamefont{Steinhardt}},
  \bibinfo{journal}{Phys. Rev.} \textbf{\bibinfo{volume}{D76}},
  \bibinfo{pages}{103501} (\bibinfo{year}{2007}{\natexlab{c}}),
  \eprint{hep-th/0702153}.

\bibitem[{\citenamefont{Tolley and Wesley}(2007)}]{Tolley:2007nq}
\bibinfo{author}{\bibfnamefont{A.~J.} \bibnamefont{Tolley}} \bibnamefont{and}
  \bibinfo{author}{\bibfnamefont{D.~H.} \bibnamefont{Wesley}},
  \bibinfo{journal}{JCAP} \textbf{\bibinfo{volume}{0705}}, \bibinfo{pages}{006}
  (\bibinfo{year}{2007}), \eprint{hep-th/0703101}.

\bibitem[{\citenamefont{Lehners and
  Steinhardt}(2009{\natexlab{a}})}]{Lehners:2008qe}
\bibinfo{author}{\bibfnamefont{J.-L.} \bibnamefont{Lehners}} \bibnamefont{and}
  \bibinfo{author}{\bibfnamefont{P.~J.} \bibnamefont{Steinhardt}},
  \bibinfo{journal}{Phys. Rev.} \textbf{\bibinfo{volume}{D79}},
  \bibinfo{pages}{063503} (\bibinfo{year}{2009}{\natexlab{a}}),
  \eprint{0812.3388}.

\bibitem[{\citenamefont{Lehners et~al.}(2009)\citenamefont{Lehners, Steinhardt,
  and Turok}}]{Lehners:2009eg}
\bibinfo{author}{\bibfnamefont{J.-L.} \bibnamefont{Lehners}},
  \bibinfo{author}{\bibfnamefont{P.~J.} \bibnamefont{Steinhardt}},
  \bibnamefont{and} \bibinfo{author}{\bibfnamefont{N.}~\bibnamefont{Turok}}
  (\bibinfo{year}{2009}), \eprint{0910.0834}.

\bibitem[{\citenamefont{Boyle et~al.}(2004)\citenamefont{Boyle, Steinhardt, and
  Turok}}]{Boyle:2003km}
\bibinfo{author}{\bibfnamefont{L.~A.} \bibnamefont{Boyle}},
  \bibinfo{author}{\bibfnamefont{P.~J.} \bibnamefont{Steinhardt}},
  \bibnamefont{and} \bibinfo{author}{\bibfnamefont{N.}~\bibnamefont{Turok}},
  \bibinfo{journal}{Phys. Rev.} \textbf{\bibinfo{volume}{D69}},
  \bibinfo{pages}{127302} (\bibinfo{year}{2004}), \eprint{hep-th/0307170}.

\bibitem[{\citenamefont{Baumann et~al.}(2007)\citenamefont{Baumann, Steinhardt,
  Takahashi, and Ichiki}}]{Baumann:2007zm}
\bibinfo{author}{\bibfnamefont{D.}~\bibnamefont{Baumann}},
  \bibinfo{author}{\bibfnamefont{P.~J.} \bibnamefont{Steinhardt}},
  \bibinfo{author}{\bibfnamefont{K.}~\bibnamefont{Takahashi}},
  \bibnamefont{and} \bibinfo{author}{\bibfnamefont{K.}~\bibnamefont{Ichiki}},
  \bibinfo{journal}{Phys. Rev.} \textbf{\bibinfo{volume}{D76}},
  \bibinfo{pages}{084019} (\bibinfo{year}{2007}), \eprint{hep-th/0703290}.

\bibitem[{\citenamefont{Khoury et~al.}(2002)\citenamefont{Khoury, Ovrut,
  Steinhardt, and Turok}}]{Khoury:2001zk}
\bibinfo{author}{\bibfnamefont{J.}~\bibnamefont{Khoury}},
  \bibinfo{author}{\bibfnamefont{B.~A.} \bibnamefont{Ovrut}},
  \bibinfo{author}{\bibfnamefont{P.~J.} \bibnamefont{Steinhardt}},
  \bibnamefont{and} \bibinfo{author}{\bibfnamefont{N.}~\bibnamefont{Turok}},
  \bibinfo{journal}{Phys. Rev.} \textbf{\bibinfo{volume}{D66}},
  \bibinfo{pages}{046005} (\bibinfo{year}{2002}), \eprint{hep-th/0109050}.

\bibitem[{\citenamefont{Creminelli et~al.}(2005)\citenamefont{Creminelli,
  Nicolis, and Zaldarriaga}}]{Creminelli:2004jg}
\bibinfo{author}{\bibfnamefont{P.}~\bibnamefont{Creminelli}},
  \bibinfo{author}{\bibfnamefont{A.}~\bibnamefont{Nicolis}}, \bibnamefont{and}
  \bibinfo{author}{\bibfnamefont{M.}~\bibnamefont{Zaldarriaga}},
  \bibinfo{journal}{Phys. Rev.} \textbf{\bibinfo{volume}{D71}},
  \bibinfo{pages}{063505} (\bibinfo{year}{2005}), \eprint{hep-th/0411270}.

\bibitem[{\citenamefont{Lyth}(2002)}]{Lyth:2001pf}
\bibinfo{author}{\bibfnamefont{D.~H.} \bibnamefont{Lyth}},
  \bibinfo{journal}{Phys. Lett.} \textbf{\bibinfo{volume}{B524}},
  \bibinfo{pages}{1} (\bibinfo{year}{2002}), \eprint{hep-ph/0106153}.

\bibitem[{\citenamefont{Tolley et~al.}(2004)\citenamefont{Tolley, Turok, and
  Steinhardt}}]{Tolley:2003nx}
\bibinfo{author}{\bibfnamefont{A.~J.} \bibnamefont{Tolley}},
  \bibinfo{author}{\bibfnamefont{N.}~\bibnamefont{Turok}}, \bibnamefont{and}
  \bibinfo{author}{\bibfnamefont{P.~J.} \bibnamefont{Steinhardt}},
  \bibinfo{journal}{Phys. Rev.} \textbf{\bibinfo{volume}{D69}},
  \bibinfo{pages}{106005} (\bibinfo{year}{2004}), \eprint{hep-th/0306109}.

\bibitem[{\citenamefont{McFadden et~al.}(2007)\citenamefont{McFadden, Turok,
  and Steinhardt}}]{McFadden:2005mq}
\bibinfo{author}{\bibfnamefont{P.~L.} \bibnamefont{McFadden}},
  \bibinfo{author}{\bibfnamefont{N.}~\bibnamefont{Turok}}, \bibnamefont{and}
  \bibinfo{author}{\bibfnamefont{P.~J.} \bibnamefont{Steinhardt}},
  \bibinfo{journal}{Phys. Rev.} \textbf{\bibinfo{volume}{D76}},
  \bibinfo{pages}{104038} (\bibinfo{year}{2007}), \eprint{hep-th/0512123}.

\bibitem[{\citenamefont{Khoury and Steinhardt}(2009)}]{Khoury:2009my}
\bibinfo{author}{\bibfnamefont{J.}~\bibnamefont{Khoury}} \bibnamefont{and}
  \bibinfo{author}{\bibfnamefont{P.~J.} \bibnamefont{Steinhardt}}
  (\bibinfo{year}{2009}), \eprint{0910.2230}.

\bibitem[{\citenamefont{Linde et~al.}(2010)\citenamefont{Linde, Mukhanov, and
  Vikman}}]{Linde:2009mc}
\bibinfo{author}{\bibfnamefont{A.}~\bibnamefont{Linde}},
  \bibinfo{author}{\bibfnamefont{V.}~\bibnamefont{Mukhanov}}, \bibnamefont{and}
  \bibinfo{author}{\bibfnamefont{A.}~\bibnamefont{Vikman}},
  \bibinfo{journal}{JCAP} \textbf{\bibinfo{volume}{1002}}, \bibinfo{pages}{006}
  (\bibinfo{year}{2010}), \eprint{0912.0944}.

\bibitem[{\citenamefont{Finelli}(2002)}]{Finelli:2002we}
\bibinfo{author}{\bibfnamefont{F.}~\bibnamefont{Finelli}},
  \bibinfo{journal}{Phys. Lett.} \textbf{\bibinfo{volume}{B545}},
  \bibinfo{pages}{1} (\bibinfo{year}{2002}), \eprint{hep-th/0206112}.

\bibitem[{\citenamefont{Notari and Riotto}(2002)}]{Notari:2002yc}
\bibinfo{author}{\bibfnamefont{A.}~\bibnamefont{Notari}} \bibnamefont{and}
  \bibinfo{author}{\bibfnamefont{A.}~\bibnamefont{Riotto}},
  \bibinfo{journal}{Nucl. Phys.} \textbf{\bibinfo{volume}{B644}},
  \bibinfo{pages}{371} (\bibinfo{year}{2002}), \eprint{hep-th/0205019}.

\bibitem[{\citenamefont{Langlois and Vernizzi}(2007)}]{Langlois:2006vv}
\bibinfo{author}{\bibfnamefont{D.}~\bibnamefont{Langlois}} \bibnamefont{and}
  \bibinfo{author}{\bibfnamefont{F.}~\bibnamefont{Vernizzi}},
  \bibinfo{journal}{JCAP} \textbf{\bibinfo{volume}{0702}}, \bibinfo{pages}{017}
  (\bibinfo{year}{2007}), \eprint{astro-ph/0610064}.

\bibitem[{\citenamefont{Khoury et~al.}(2003)\citenamefont{Khoury, Steinhardt,
  and Turok}}]{Khoury:2003vb}
\bibinfo{author}{\bibfnamefont{J.}~\bibnamefont{Khoury}},
  \bibinfo{author}{\bibfnamefont{P.~J.} \bibnamefont{Steinhardt}},
  \bibnamefont{and} \bibinfo{author}{\bibfnamefont{N.}~\bibnamefont{Turok}},
  \bibinfo{journal}{Phys. Rev. Lett.} \textbf{\bibinfo{volume}{91}},
  \bibinfo{pages}{161301} (\bibinfo{year}{2003}), \eprint{astro-ph/0302012}.

\bibitem[{\citenamefont{Lehners and Turok}(2008)}]{Lehners:2007nb}
\bibinfo{author}{\bibfnamefont{J.-L.} \bibnamefont{Lehners}} \bibnamefont{and}
  \bibinfo{author}{\bibfnamefont{N.}~\bibnamefont{Turok}},
  \bibinfo{journal}{Phys. Rev.} \textbf{\bibinfo{volume}{D77}},
  \bibinfo{pages}{023516} (\bibinfo{year}{2008}), \eprint{arXiv:0708.0743
  [hep-th]}.

\bibitem[{\citenamefont{Lehners and
  Steinhardt}(2008{\natexlab{a}})}]{Lehners:2008my}
\bibinfo{author}{\bibfnamefont{J.-L.} \bibnamefont{Lehners}} \bibnamefont{and}
  \bibinfo{author}{\bibfnamefont{P.~J.} \bibnamefont{Steinhardt}},
  \bibinfo{journal}{Phys. Rev.} \textbf{\bibinfo{volume}{D78}},
  \bibinfo{pages}{023506} (\bibinfo{year}{2008}{\natexlab{a}}),
  \eprint{0804.1293}.

\bibitem[{\citenamefont{Battefeld}(2008)}]{Battefeld:2007st}
\bibinfo{author}{\bibfnamefont{T.}~\bibnamefont{Battefeld}},
  \bibinfo{journal}{Phys. Rev.} \textbf{\bibinfo{volume}{D77}},
  \bibinfo{pages}{063503} (\bibinfo{year}{2008}), \eprint{0710.2540}.

\bibitem[{\citenamefont{Byrnes et~al.}(2006)\citenamefont{Byrnes, Sasaki, and
  Wands}}]{Byrnes:2006vq}
\bibinfo{author}{\bibfnamefont{C.~T.} \bibnamefont{Byrnes}},
  \bibinfo{author}{\bibfnamefont{M.}~\bibnamefont{Sasaki}}, \bibnamefont{and}
  \bibinfo{author}{\bibfnamefont{D.}~\bibnamefont{Wands}},
  \bibinfo{journal}{Phys. Rev.} \textbf{\bibinfo{volume}{D74}},
  \bibinfo{pages}{123519} (\bibinfo{year}{2006}), \eprint{astro-ph/0611075}.

\bibitem[{\citenamefont{Babich et~al.}(2004)\citenamefont{Babich, Creminelli,
  and Zaldarriaga}}]{Babich:2004gb}
\bibinfo{author}{\bibfnamefont{D.}~\bibnamefont{Babich}},
  \bibinfo{author}{\bibfnamefont{P.}~\bibnamefont{Creminelli}},
  \bibnamefont{and}
  \bibinfo{author}{\bibfnamefont{M.}~\bibnamefont{Zaldarriaga}},
  \bibinfo{journal}{JCAP} \textbf{\bibinfo{volume}{0408}}, \bibinfo{pages}{009}
  (\bibinfo{year}{2004}), \eprint{astro-ph/0405356}.

\bibitem[{\citenamefont{Maldacena}(2003)}]{Maldacena:2002vr}
\bibinfo{author}{\bibfnamefont{J.~M.} \bibnamefont{Maldacena}},
  \bibinfo{journal}{JHEP} \textbf{\bibinfo{volume}{05}}, \bibinfo{pages}{013}
  (\bibinfo{year}{2003}), \eprint{astro-ph/0210603}.

\bibitem[{\citenamefont{Koyama et~al.}(2007{\natexlab{b}})\citenamefont{Koyama,
  Mizuno, Vernizzi, and Wands}}]{Koyama:2007if}
\bibinfo{author}{\bibfnamefont{K.}~\bibnamefont{Koyama}},
  \bibinfo{author}{\bibfnamefont{S.}~\bibnamefont{Mizuno}},
  \bibinfo{author}{\bibfnamefont{F.}~\bibnamefont{Vernizzi}}, \bibnamefont{and}
  \bibinfo{author}{\bibfnamefont{D.}~\bibnamefont{Wands}},
  \bibinfo{journal}{JCAP} \textbf{\bibinfo{volume}{0711}}, \bibinfo{pages}{024}
  (\bibinfo{year}{2007}{\natexlab{b}}), \eprint{0708.4321}.

\bibitem[{\citenamefont{Lehners and Renaux-Petel}(2009)}]{Lehners:2009ja}
\bibinfo{author}{\bibfnamefont{J.-L.} \bibnamefont{Lehners}} \bibnamefont{and}
  \bibinfo{author}{\bibfnamefont{S.}~\bibnamefont{Renaux-Petel}},
  \bibinfo{journal}{Phys. Rev.} \textbf{\bibinfo{volume}{D80}},
  \bibinfo{pages}{063503} (\bibinfo{year}{2009}), \eprint{0906.0530}.

\bibitem[{\citenamefont{Starobinsky}(1985)}]{Starobinsky:1986fxa}
\bibinfo{author}{\bibfnamefont{A.~A.} \bibnamefont{Starobinsky}},
  \bibinfo{journal}{JETP Lett.} \textbf{\bibinfo{volume}{42}},
  \bibinfo{pages}{152} (\bibinfo{year}{1985}).

\bibitem[{\citenamefont{Sasaki and Stewart}(1996)}]{Sasaki:1995aw}
\bibinfo{author}{\bibfnamefont{M.}~\bibnamefont{Sasaki}} \bibnamefont{and}
  \bibinfo{author}{\bibfnamefont{E.~D.} \bibnamefont{Stewart}},
  \bibinfo{journal}{Prog. Theor. Phys.} \textbf{\bibinfo{volume}{95}},
  \bibinfo{pages}{71} (\bibinfo{year}{1996}), \eprint{astro-ph/9507001}.

\bibitem[{\citenamefont{Lyth et~al.}(2005)\citenamefont{Lyth, Malik, and
  Sasaki}}]{Lyth:2004gb}
\bibinfo{author}{\bibfnamefont{D.~H.} \bibnamefont{Lyth}},
  \bibinfo{author}{\bibfnamefont{K.~A.} \bibnamefont{Malik}}, \bibnamefont{and}
  \bibinfo{author}{\bibfnamefont{M.}~\bibnamefont{Sasaki}},
  \bibinfo{journal}{JCAP} \textbf{\bibinfo{volume}{0505}}, \bibinfo{pages}{004}
  (\bibinfo{year}{2005}), \eprint{astro-ph/0411220}.

\bibitem[{\citenamefont{Buchbinder et~al.}(2008)\citenamefont{Buchbinder,
  Khoury, and Ovrut}}]{Buchbinder:2007at}
\bibinfo{author}{\bibfnamefont{E.~I.} \bibnamefont{Buchbinder}},
  \bibinfo{author}{\bibfnamefont{J.}~\bibnamefont{Khoury}}, \bibnamefont{and}
  \bibinfo{author}{\bibfnamefont{B.~A.} \bibnamefont{Ovrut}},
  \bibinfo{journal}{Phys. Rev. Lett.} \textbf{\bibinfo{volume}{100}},
  \bibinfo{pages}{171302} (\bibinfo{year}{2008}), \eprint{0710.5172}.

\bibitem[{\citenamefont{Lehners and
  Steinhardt}(2009{\natexlab{b}})}]{Lehners:2009qu}
\bibinfo{author}{\bibfnamefont{J.-L.} \bibnamefont{Lehners}} \bibnamefont{and}
  \bibinfo{author}{\bibfnamefont{P.~J.} \bibnamefont{Steinhardt}},
  \bibinfo{journal}{Phys. Rev.} \textbf{\bibinfo{volume}{D80}},
  \bibinfo{pages}{103520} (\bibinfo{year}{2009}{\natexlab{b}}),
  \eprint{0909.2558}.

\bibitem[{\citenamefont{Lehners and
  Steinhardt}(2008{\natexlab{b}})}]{Lehners:2007wc}
\bibinfo{author}{\bibfnamefont{J.-L.} \bibnamefont{Lehners}} \bibnamefont{and}
  \bibinfo{author}{\bibfnamefont{P.~J.} \bibnamefont{Steinhardt}},
  \bibinfo{journal}{Phys. Rev.} \textbf{\bibinfo{volume}{D77}},
  \bibinfo{pages}{063533} (\bibinfo{year}{2008}{\natexlab{b}}),
  \eprint{0712.3779}.

\bibitem[{\citenamefont{Smith et~al.}(2009)\citenamefont{Smith, Senatore, and
  Zaldarriaga}}]{Smith:2009jr}
\bibinfo{author}{\bibfnamefont{K.~M.} \bibnamefont{Smith}},
  \bibinfo{author}{\bibfnamefont{L.}~\bibnamefont{Senatore}}, \bibnamefont{and}
  \bibinfo{author}{\bibfnamefont{M.}~\bibnamefont{Zaldarriaga}},
  \bibinfo{journal}{JCAP} \textbf{\bibinfo{volume}{0909}}, \bibinfo{pages}{006}
  (\bibinfo{year}{2009}), \eprint{0901.2572}.

\bibitem[{\citenamefont{Khoury and Piazza}(2009)}]{Khoury:2008wj}
\bibinfo{author}{\bibfnamefont{J.}~\bibnamefont{Khoury}} \bibnamefont{and}
  \bibinfo{author}{\bibfnamefont{F.}~\bibnamefont{Piazza}},
  \bibinfo{journal}{JCAP} \textbf{\bibinfo{volume}{0907}}, \bibinfo{pages}{026}
  (\bibinfo{year}{2009}), \eprint{0811.3633}.

\bibitem[{\citenamefont{Seery and Lidsey}(2007)}]{Seery:2006js}
\bibinfo{author}{\bibfnamefont{D.}~\bibnamefont{Seery}} \bibnamefont{and}
  \bibinfo{author}{\bibfnamefont{J.~E.} \bibnamefont{Lidsey}},
  \bibinfo{journal}{JCAP} \textbf{\bibinfo{volume}{0701}}, \bibinfo{pages}{008}
  (\bibinfo{year}{2007}), \eprint{astro-ph/0611034}.

\bibitem[{\citenamefont{Bartolo et~al.}(2004)\citenamefont{Bartolo, Komatsu,
  Matarrese, and Riotto}}]{Bartolo:2004if}
\bibinfo{author}{\bibfnamefont{N.}~\bibnamefont{Bartolo}},
  \bibinfo{author}{\bibfnamefont{E.}~\bibnamefont{Komatsu}},
  \bibinfo{author}{\bibfnamefont{S.}~\bibnamefont{Matarrese}},
  \bibnamefont{and} \bibinfo{author}{\bibfnamefont{A.}~\bibnamefont{Riotto}},
  \bibinfo{journal}{Phys. Rept.} \textbf{\bibinfo{volume}{402}},
  \bibinfo{pages}{103} (\bibinfo{year}{2004}), \eprint{astro-ph/0406398}.

\end{thebibliography}

\end{document}